%% file: main.tex
\documentclass{natureprintstyle}

\usepackage{aas_macros}
\usepackage{amsmath}
\usepackage{amssymb}
\usepackage[colorlinks,citecolor=blue,linkcolor=cyan]{hyperref}
\usepackage{graphicx}
\usepackage{multirow}

\newcommand{\methref}[1]{\hyperref[#1]{Methods}}
\newcommand{\edfref}[1]{\hyperref[#1]{Extended Data Figure~\ref{#1}}}
\newcommand{\edtref}[1]{\hyperref[#1]{Extended Data Table~\ref{#1}}}
\newcommand{\sisecref}[1]{\hyperref[#1]{SI Section~\ref{#1}}}

\usepackage{scalerel,tikz}
\usetikzlibrary{svg.path}
\definecolor{orcidlogocol}{HTML}{A6CE39}
\tikzset{orcidlogo/.pic={
 \fill[orcidlogocol] svg{M256,128c0,70.7-57.3,128-128,128C57.3,256,0,198.7,0,128C0,57.3,57.3,0,128,0C198.7,0,256,57.3,256,128z};
 \fill[white] svg{M86.3,186.2H70.9V79.1h15.4v48.4V186.2z}
 svg{M108.9,79.1h41.6c39.6,0,57,28.3,57,53.6c0,27.5-21.5,53.6-56.8,53.6h-41.8V79.1z M124.3,172.4h24.5c34.9,0,42.9-26.5,42.9-39.7c0-21.5-13.7-39.7-43.7-39.7h-23.7V172.4z}
 svg{M88.7,56.8c0,5.5-4.5,10.1-10.1,10.1c-5.6,0-10.1-4.6-10.1-10.1c0-5.6,4.5-10.1,10.1-10.1C84.2,46.7,88.7,51.3,88.7,56.8z};
}}
\newcommand\orcidicon[1]{\href{https://orcid.org/#1}{\mbox{\scalerel*{
\begin{tikzpicture}[yscale=-1,transform shape]
\pic{orcidlogo};
\end{tikzpicture}
}{|}}}}

\title{TeV $\gamma$-ray emission near globular cluster Terzan 5 as a probe of cosmic ray transport}

\author{Mark R.~Krumholz$^{\orcidicon{0000-0003-3893-854X}1,2}$, Roland M.~Crocker$^2$, Arash Bahramian$^3$, \& Pol Bordas$^4$}

\begin{document}

\maketitle

\begin{affiliations}
	\item Research School of Astronomy and Astrophysics, Australian National University, Canberra 2611, A.C.T., Australia
	\item ARC Centre of Excellence for Astronomy in Three Dimensions (ASTRO-3D), Canberra 2611, A.C.T., Australia
	\item International Centre for Radio Astronomy Research - Curtin University, 1 Turner Ave, Bentley, WA 6102, Australia
	\item Departament de Física Qu\`{a}ntica i Astrofísica, Institut de Ci\`{e}ncies del Cosmos, Universitat de Barcelona, IEEC-UB, Martí i Franqu\`{e}s 1, E-08028 Barcelona, Spain\\
\end{affiliations}

\begin{abstract}
Cosmic rays travelling through interstellar space have their propagation directions repeatedly scattered by fluctuating interstellar magnetic fields. The nature of this scattering is a major unsolved problem in astrophysics, one that has resisted solution largely due to a lack of direct observational constraints on the scattering rate. Here we show that very high-energy $\gamma$-ray emission from the globular cluster Terzan 5, which has unexpectedly been found to be displaced from the cluster, presents a direct probe of this process. We show that this displacement is naturally explained by cosmic rays accelerated in the bow shock around the cluster propagating a finite distance before scattering processes re-orient enough of them towards Earth to produce a detectable $\gamma$-ray signal. The angular distance between the cluster and the signal places tight constraints on the scattering rate, which we show are consistent with a model whereby scattering is primarily due to excitation of magnetic waves by the cosmic rays themselves. The analysis method we develop here will make it possible to use sources with similarly displaced non-thermal X-ray and TeV $\gamma$-ray signals as direct probes of cosmic ray scattering across a range of Galactic environments.
\end{abstract}

Cosmic rays (CRs) are charged non-thermal particles accelerated to high velocities by a variety of collisionless magnetic processes. These particles and their interactions with interstellar gas, radiation, and magnetic fields are responsible for most of the diffuse $\gamma$-ray emission we see on the sky. This diffuse signal is brightest around sources that accelerate CRs, most prominently supernova remnants\cite{Helder2012}, but also young star clusters\cite{Ackermann16a}, old star clusters\cite{Song2021}, and dwarf galaxies\cite{Crocker2022}. In the latter two cases, CRs are believed to be predominantly $e^\pm$ pairs (CRe) that are accelerated in the magnetospheres of millisecond pulsars (MSPs) and then injected into the interstellar medium\cite{Sudoh2021, Gautam2022}.

\begin{figure}
	\includegraphics[width=\columnwidth]{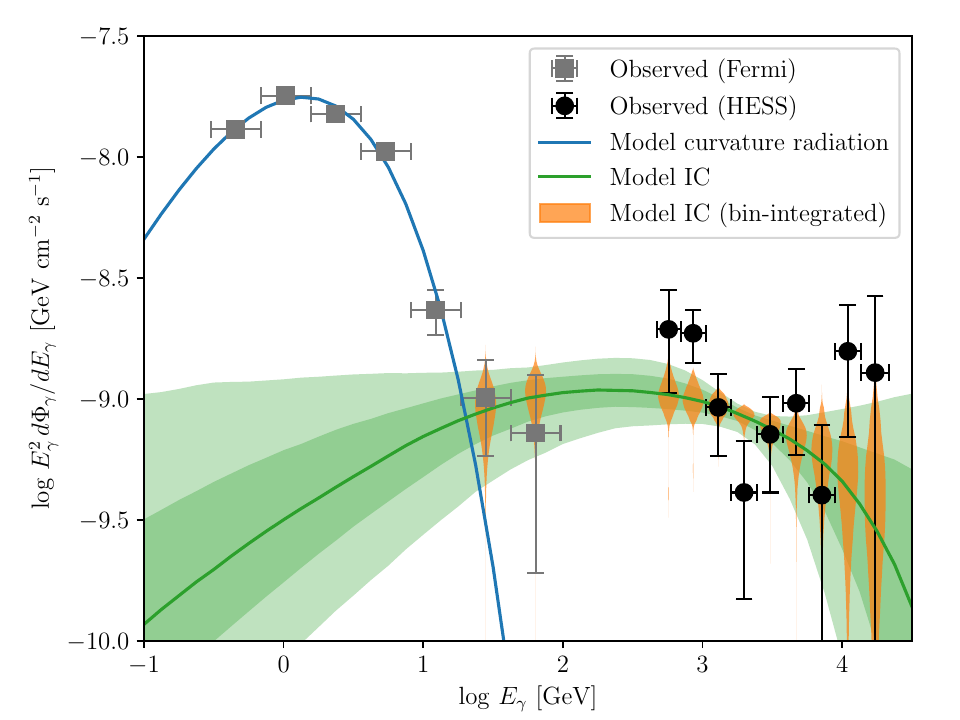}
	\caption{ 
	\textbf{The measured $\gamma$-ray spectrum of Terzan 5 together with model fits.} Black and grey points with error bars show measured values of the photon flux $d\Phi_\gamma/dE_\gamma$, as a function of photon energy $E_\gamma$; we compensate the photon flux by a factor of $E_\gamma^2$ for plotting convenience. Squares show measurements from \textit{Fermi}, circles from HESS; horizontal error bars show the energy range over which the flux is measured. The blue solid line is a fit to the $E_\gamma<20$ GeV measurements using an exponentially-truncated powerlaw functional form, as expected for curvature radiation from MSPs; this clearly fails to reproduce the higher-energy emission. The green solid line is the median inverse Compton spectrum predicted by the model presented here; the shaded bands around it show the 68\% and 95\% confidence interval around this median. Orange violin plots show the PDF of bin-averaged photon fluxes computed from the model at each of the energy bins used to fit the model to the data; see \methref{sec:Methods} for details.}
\label{fig:spectrum}
\end{figure}

The brightest globular cluster (GC) in $\gamma$-rays\cite{Song2021}, and the only GC thus far detected (by the High Energy Stereoscopic System, HESS) at $\gamma$-ray energies $\gtrsim 100$ GeV, is Terzan 5\cite{Abramowski2011}; we show the measured spectrum in \autoref{fig:spectrum}. While Terzan 5's emission at $\lesssim 10$ GeV energies is likely driven by curvature radiation by CRe with energies $\lesssim 3$ TeV in MSP magnetospheres, a fit to these data using the truncated powerlaw distribution expected for this emission mechanism is clearly unable to explain the observed emission at higher energies, which reaches to $\sim$20 TeV. Instead this emission is likely a result of inverse Compton (IC) scattering by CRe that escape from the pulsars and are then re-accelerated at the shock where the combined MSP winds encounter the interstellar medium (ISM), increasing the maximum CR energy from the $\approx 3$ TeV expected based on the curvature radiation spectrum to the tens of TeV required to explain the higher-energy emission\cite{Bednarek2014, Bednarek2016}; such acceleration is routinely observed around the bow shocks of single MSPs\cite{Bykov2017}.

Intriguingly, however, while emission from Terzan 5 at lower $\gamma$-ray energies is well-centred on the cluster's optical position\cite{Ndiyavala2019}, the HESS signal is displaced by an angular distance of $\Delta\theta = 4'\!\!.0 \pm 28''\!\!\!.5$, corresponding to a projected physical offset of $7.7\pm 0.9$ pc at the $d_\mathrm{Ter5} = 6.62$ kpc distance\cite{Baumgardt21a} of the cluster; this offset is much larger than the $\approx 0.29$ pc core radius of the cluster's red giant star distribution\cite{Lanzoni2010} or its even more concentrated MSP population with a $\approx 0.18$ pc core radius\cite{Prager2017}. (These two lengths, and all others appearing in the remainder of this paper that are derived from angular measurements, have been homogenised to use the 6.62 kpc distance to Terzan 5 determined by ref.~\cite{Baumgardt21a}.) The reason for this displacement remains unknown. It is far larger than any plausible standoff distance between the cluster and the shock, which is likely $< 1$ pc\cite{Bednarek2014}. While CRs accelerated in the shock may plausibly be advected $\sim 10$ pc before cooling\cite{Bednarek2014}, this would not explain why the maximum of emission does not coincide with the acceleration location, where the CR energy density is largest because no cooling has yet taken place, and close to where the density of photons available for IC scattering also peaks due to the cluster's own radiation field.

However, the displacement is naturally explained if we drop the implicit assumption that the CRs are everywhere pitch angle-isotropised; in fact, the situation of `colliding shock flows' realised inside the Terzan 5 bow shock is expected to generate a highly anisotropic pitch angle distribution\cite{Bykov2017}. We illustrate the basic geometry in \autoref{fig:Terzan5_schematic} and \edfref{fig:T5geometry}: the cluster moves through the ISM at $\mathcal{O}( 100$ km s$^{-1})$, and its motion generates {both a bow shock where the collective pulsar winds encounter the ISM and a magnetotail behind the cluster where the magnetic fields of the ISM are swept up. Compression at the bow shock produces a field strength that is roughly such that the post-shock Alfv\'en speed is comparable to the cluster speed, implying $B \approx v_\mathrm{c} \sqrt{4\pi \mu_\mathrm{H} n_\mathrm{H}} = 54 (v_\mathrm{c}/100\;\mathrm{km s}^{-1}) (n_\mathrm{H}/1\;\mathrm{cm}^{-3})^{1/2}$ $\mu$G; the normalisation chosen for $n_\mathrm{H}$ is a reasonable if rough estimate of the ISM number density in the vicinity of Terzan 5 given its proximity to the Galactic plane. While simulations of pulsar wind bow shocks suggest that the shock and magnetotail will be smooth on large scales as long as the MSP winds are strongly magnetised\cite{Barkov19a, Olmi19a, Olmi2023}, near the head of the shock the large flux of non-thermal electrons in the pulsar wind drives instabilities on small ($\sim 100$ AU) scales, resulting in turbulence that accelerates particles. As the particles reach higher energies their mean free path increases, and those with small pitch angles relative to the large-scale smooth field, whose motion is therefore predominantly along that field, begin to be able to travel far enough to escape from the accelerator region into the magnetotail. (We estimate the size of this region, and show that its TeV $\gamma$-ray signal is too weak to be seen by HESS, in the Supplementary Information.) Because the CRe escaping the acceleration zone have a narrow pitch angle distribution and are ultrarelativistic (Lorentz factors $\gamma_e \gtrsim 10^6$), their emission is beamed into an extremely narrow cone, and thus is initially invisible to us if the tail is not sufficiently aligned towards our line-of-sight. As the CRe travel down the magnetic field trailing the cluster, though, magnetic fluctuations scatter their pitch angles, broadening the distribution. We begin to see IC emission once the distribution is broad enough to include a significant population of CRe whose pitch angles are such that their radiation beam passes over us at some point during their gyroscopic motion. However, by the time this occurs, the CRe have moved a finite distance down the magnetotail, resulting in the observed displacement between the emission and the cluster.

\begin{figure}
        \includegraphics[width=\columnwidth]{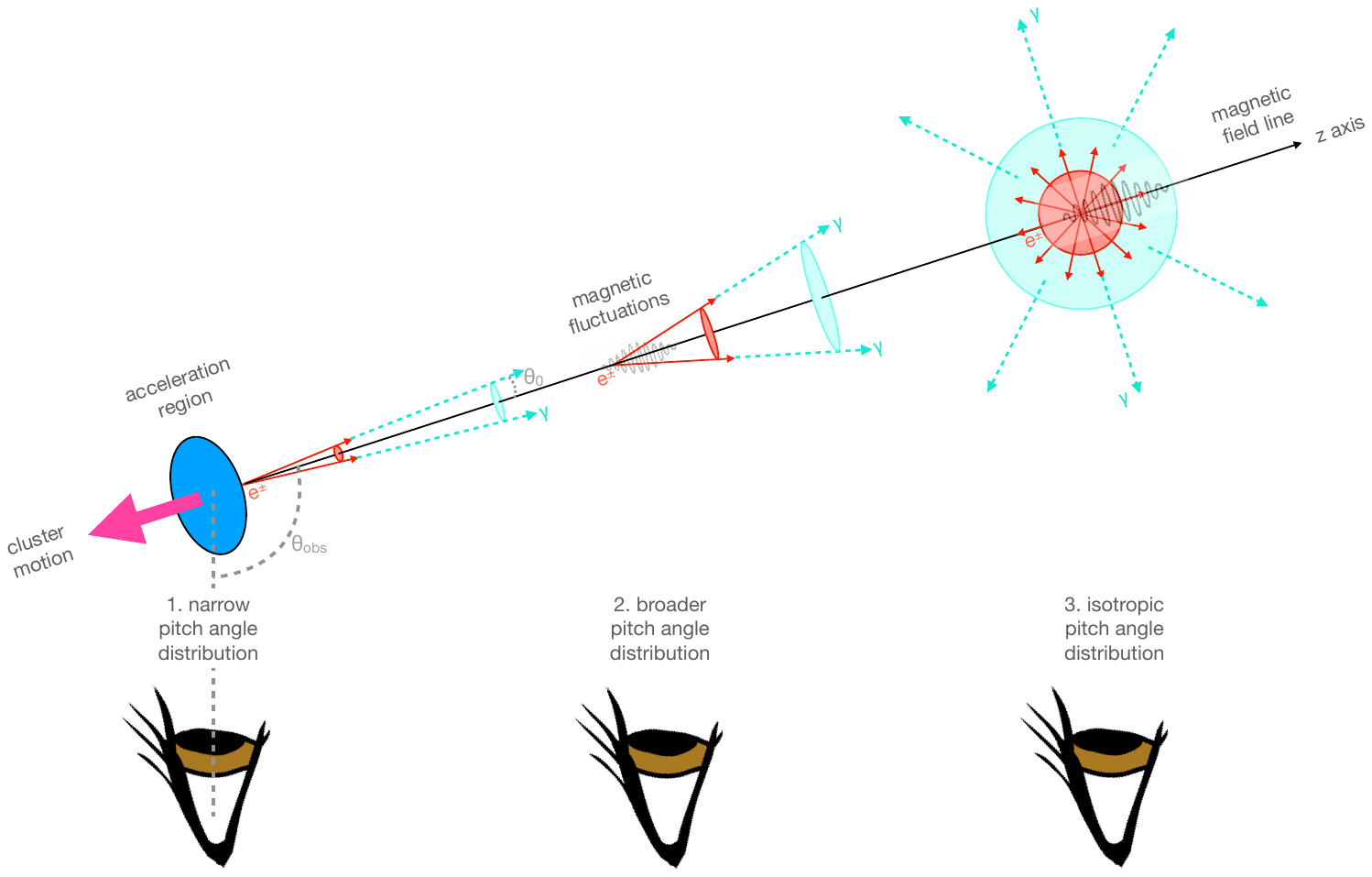}
\caption{ 
{\bf Schematic illustration of the mechanism underlying the displaced TeV $\gamma$-ray signal of Terzan 5}. The supersonic motion of the globular cluster through the ISM (to the left here) generates a trailing magnetotail. Cosmic ray $e^\pm$ escaping from MSPs in the cluster are reaccelerated at the shock between the MSP wind and the ISM, and sufficiently high energy CR $e^\pm$s with pitch angles $\theta < \theta_0 < \pi/2$ relative to the field escape down the magnetotail. Since our line of sight is not within the narrow range of pitch angles of the CRs initially injected down the tail, the radiation they emit is invisible to us. However, as CRs propagate down the field they random walk in pitch angle due to magnetic fluctuations, which drives the pitch angle distribution towards isotropy and eventually renders the CRs visible to us.
}
\label{fig:Terzan5_schematic}
\end{figure}

To model this process we consider a population of ultrarelativistic CRe injected at the origin at a rate $\dot{N}$ with a momentum distribution $dn/dp \propto p^{k_p} \exp(-p/p_\mathrm{cut})$ for $p>p_\mathrm{min}$ with some range of pitch angles $(0,\theta_0)$; for numerical purposes we adopt $p_\mathrm{min} = 1$ GeV$/c$, but this choice has no practical effect on our results as long as it is small enough that CRs of momentum $p_\mathrm{min}$ do not contribute to the observed emission. There is a background magnetic field of constant strength aligned parallel to the $z$ direction, which wraps around the cluster at the origin. We average over the gyroscopic motion, so that a CR with a cosine pitch angle $\mu = \cos\theta$ simply moves down the field at a velocity $v_z = \mu c$, and the wrapping of the field generates a reflecting boundary condition at $z=0$. The CRe diffuse in pitch angle with a constant diffusion coefficient $K_\mu$. For the characteristic magnetic field strength estimated above and the radiation field estimates that we shall describe below, synchrotron emission is much stronger than any other loss mechanism except for $\lesssim 100$ GeV CRe located within the small cluster core radius ($r \approx 0.3$ pc), where IC losses become competitive; since synchrotron emission dominates over both the energy range and the locations where we observe the HESS signal, we neglect other loss mechanisms. Synchrotron emission causes CRs to lose momentum at a rate $dp/dt = -(p/m_e c)^2 (1-\mu^2) t_{c,0}^{-1}$, where $t_{c,0} = 4\pi m_e c / \sigma_\mathrm{T} B^2$ is the synchrotron cooling timescale at momentum $p=m_e c$ and $\sigma_\mathrm{T}$ is the Thomson cross section. In principle our model could also include parameters to describe covariance between the spectral index $k_p$ and the maximum opening angle $\theta_0$, since higher momentum particles have longer mean free paths and thus can escape the accelerator region over a wider range of pitch angles. However, current models do not predict a quantitative form for this covariance, and thus we choose the simplest option of neglecting it. We will find below that in our final results for best-fitting parameters there is no noticeable covariance between  $k_p$ and $\theta_0$ in the posterior probability distribution, which suggests that introducing such a covariance into our model would not have large effects.

For the system we have described, the CRe distribution function $f(z, p, \mu)$ in position, momentum, and pitch angle obeys the evolution equation
\begin{eqnarray}
    \frac{\partial f}{\partial t} & = &
    -\mu c \frac{\partial f}{\partial z} + 
    \frac{\partial}{\partial \mu}\left[\left(1-\mu^2\right)K_\mu\frac{\partial f}{\partial\mu}\right] + {}
    \nonumber \\
    & &
    \frac{m_e c}{t_{c,0}}\frac{1}{p^2}\frac{\partial}{\partial p}\left[p^2 \left(1-\mu^2\right)\left(\frac{p}{m_e c}\right)^2 f\right] + {}
    \nonumber \\
    & & \dot{N}\frac{dn}{dp} \delta(z) \Theta(\mu-\mu_0),
    \label{eq:fpe}
\end{eqnarray}
where $\Theta(x)$ is the Heaviside step function; we have a zero flux boundary condition at $z=0$ and open boundaries as $z\to\infty$. We show in \methref{sec:Methods} that this equation admits a self-similar steady state solution that depends on pitch angle, dimensionless position $\zeta = (K_\mu/c) z$, and dimensionless momentum $q = p/p_\mathrm{eq}$, where $p_\mathrm{eq} = K_\mu t_{c,0} m_e c$ is the momentum for which the synchrotron loss timescale $p/|dp/dt|$ and pitch angle scattering timescale $1/K_\mu$ are equal. We solve \autoref{eq:fpe} numerically to find this steady-state solution using the CR transport code \textsc{criptic}\cite{Krumholz22a} (see \methref{sec:Methods}).

We illustrate some sample solutions to this problem in \autoref{fig:fpe_sample_solution}. The critical property that is apparent from these solutions is that, at CR momenta $q \sim 1$ and $\mu > \mu_0 \equiv \cos\theta_0$, the maximum of the distribution is displaced to positive $\zeta$, while for $q \ll 1$ the distribution is much broader, peaks near $\zeta = 0$, and is nearly uniform in pitch angle. The physical reason for this is simply that the synchrotron cooling time scales as $1/q$, so CRs that initially have $q \gtrsim 1$ cool before they have time to isotropise in pitch angle and thereafter in space, leaving a distribution that reflects the initial asymmetry present at injection, while for $q \ll 1$ isotropisation happens faster than cooling, leaving a solution with no significant pitch angle structure or displacement. This phenomenon offers a natural explanation for the displaced high-energy $\gamma$-ray emission. By contrast, we show in the Supplementary Information that two alternate scenarios for the TeV displacement fail to reproduce the observations in some critical aspects.

\begin{figure}
        \includegraphics[width=\columnwidth]{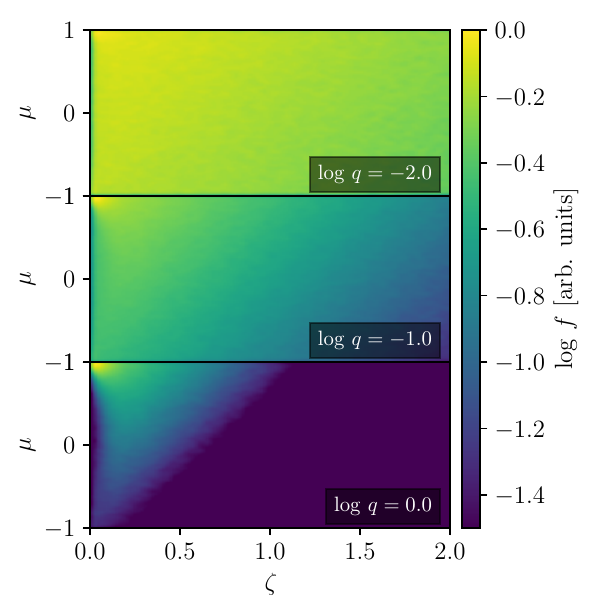}
\caption{ 
\textbf{Sample steady-state CR position-pitch angle joint distributions.} The three panels show the steady-state CR distribution function $f(\zeta,q,\mu)$ obtained by solving \autoref{eq:fpe} at three example momenta, $q = 0.01$, $0.1$, and $1.0$ (top to bottom), for the example parameters $k_p = -1.5$, $q_\mathrm{cut} = 10^4$, and $\mu_0 = \cos\theta_0 = 0.9$. Colours indicate the density of CRs at each point in position-pitch angle space, and have been normalised to have a maximum of unity in each panel. Note how for $q\approx 1$ and intermediate pitch angles the distribution has a maximum displaced to positive $\zeta$, and well away from $\zeta = 0$. We do not show cases with $q\gg 1$ because their pitch angle-anisotropy is even more extreme than for the $q=1$ case, and as a result such CRs will not be visible to observers at all unless they are located at $\mu > \mu_0$, i.e., within the range of pitch angles present at injection.
}
\label{fig:fpe_sample_solution}
\end{figure}

To evaluate this scenario quantitatively, we estimate the radiation field near Terzan 5 and compute the expected spatially-integrated spectrum $d\Phi_\gamma/dE_\gamma$ and the energy-integrated emission as a function of angular displacement from Terzan 5, $d\Phi_\gamma/d\theta$, as a function of seven parameters (see \methref{sec:Methods} for details). Three of these -- $\mu_0$, $k_p$, and $q_\mathrm{cut} = p_\mathrm{cut}/p_\mathrm{eq}$ -- determine the dimensionless CRe distribution function at injection, and three more -- $K_\mu$, $p_\mathrm{eq}$, and $\dot{N}$ -- determine the mapping from dimensonless to dimensional parameters and thus specify the distribution function in physical units. The final parameter, $\mu_\mathrm{obs}$, is the cosine of the angle between the magnetic field and the vector pointing from Terzan 5 to the Sun, and thus determines which CRe pitch angles produce radiation that we can observe. Intuitively, to explain why the IC signal is displaced $\sim 10$ pc from the cluster we must have $K_\mu \sim c/10\;\mathrm{pc}\sim 10^{-9}$ s$^{-1}$, and to explain why the TeV signal is displaced and fairly narrow, we must have comparable cooling and isotropisation times at $p_\mathrm{eq}\sim 1-10\, \mathrm{TeV}/c$. Finally, since the observed photons have a spectral index of $-2$ or slightly harder (\autoref{fig:spectrum}), and the CRs producing it are cooled, the injection spectrum must have $k_p > -2$. We confirm these intuitions, and give more precise bounds on various parameters, using a Markov Chain Monte Carlo (MCMC; see \methref{sec:Methods}) to determine which combinations of parameters successfully reproduce both the observed spectrum and the spatial distribution of emission. We find a range of parameters that are consistent with the data, with reduced $\chi^2$ values close to unity; \edtref{tab:posteriors} gives quantitative values. We compare our model-predicted spectra to the observations in \autoref{fig:spectrum}, and our model-predicted angular distribution to the HESS observations in \autoref{fig:profile}. These figures demonstrate that our models simultaneously reproduce the displaced emission peak and angular width of the HESS signal and spatially-integrated spectrum. 

\begin{figure}
    \centering
        \includegraphics[width=\columnwidth]{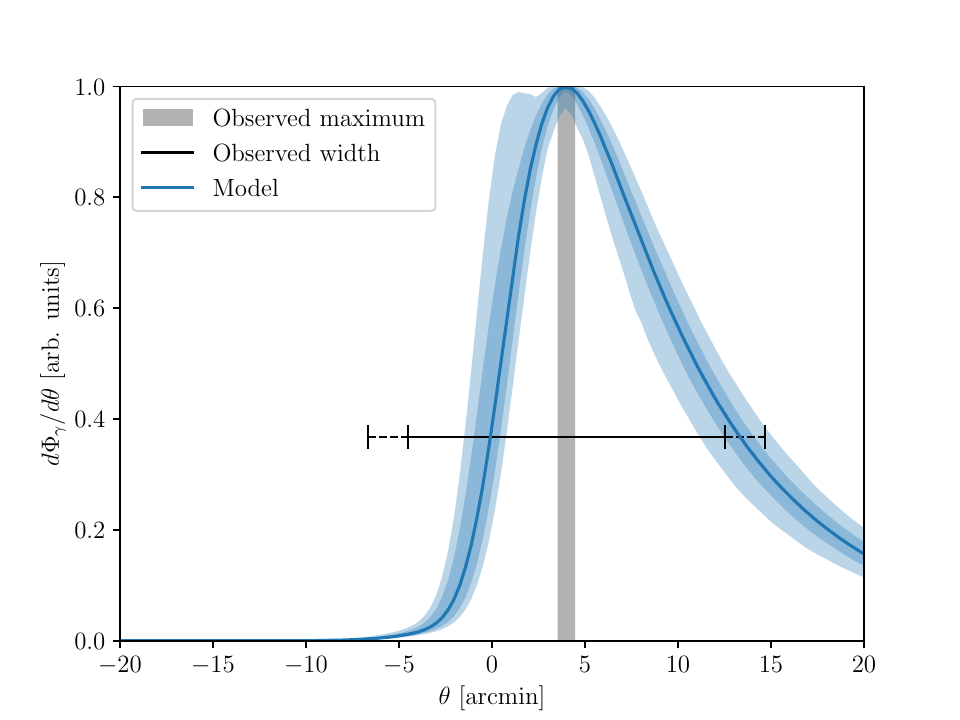}
        \caption{
        \textbf{Comparison between observed and model-predicted angular distribution of emission.} The blue line shows the median best-fitting model prediction of the angular profile of emission observable by HESS, $d\Phi_\gamma/d\theta$, as a function of angular displacement from the cluster $\theta$; shaded bands show the 68\% and 95\% confidence intervals, and all angular profiles have been normalised to a peak value of unity. The vertical grey strip shows the estimated position of peak emission $\theta_1 = 4'\!\!.0 \pm 28''\!\!\!.5$ observed by HESS, while the black horizontal line with error bars shows the angular dispersion $\theta_2 = 9'\!\!.6 \pm 36''$ of the HESS emission estimated from a Gaussian fit to the image \cite{Abramowski2011}. 
        \label{fig:profile}
        }
\end{figure}

From the MCMC results we are able to infer marginal posterior distributions for each of the model parameters, and for the total CRe kinetic luminosity and magnetic field strength, which can be derived from them (see \methref{sec:Methods}). We report these marginal posteriors in \edtref{tab:posteriors}, and we plot the full joint posterior distributions in \edfref{fig:corner}. We find a pitch angle scattering rate $K_\mu = 1.1^{+1.5}_{-0.9}\times 10^{-10}$ s$^{-1}$ (95\% confidence interval), corresponding to a spatial diffusion coefficient\cite{Zweibel2017} $K_{x} = c^2/ 6K_\mu = 1.4_{-0.8}^{+5.5}\times 10^{30}$ cm$^2$ s$^{-1}$. The spatial diffusion coefficient is similar to values typically estimated at TeV energies elsewhere in the Milky Way disc\cite{Amato18a, De-La-Torre-Luque21a}, but larger than the reduced values sometimes inferred around other TeV sources\cite{Abeysekara17a, Lopez-Coto22a}. Given the various ways in which Terzan 5's magnetotail differs from both mean Galactic conditions and those around other TeV sources (see Supplementary Information for details), we should not expect precise numerical agreement with either, however, and we provide a more detailed comparison between our inferred value of $K_\mu$ and theoretical expectations below. Our estimate for $\mu_\mathrm{obs}$ implies a direction for the magnetotail shown in \edfref{fig:T5geometry}; unfortunately we do no have a good expectation for in which direction the tail should point, because the cluster is close to the Galactic Bar, where the velocity of the ISM it not purely circular and likely contains a substantial radial component, so that its orientation -- and thus the expected direction of the magnetotail -- is unknown\cite{Tress20a}. Other fit parameters are consistent with theoretical expectations and other constraints: the magnetic field strength $B = 110^{+80}_{-40}$ $\mu$G matches our rough estimates based on pressure balance in the cluster bow shock; we recover the hard spectral index, $k_p = -1.5^{+1.3}_{-0.6}$, expected for colliding shock flow acceleration\cite{Bykov2017}, and our total CRe luminosity $L = 2.5_{-2.1}^{+7.9}\times 10^{37}$ erg s$^{-1}$ is consistent with estimates based on the GeV-band curvature radiation\cite{Abdo2013,Bednarek2014,Crocker2022} and the pulsar spin-down power\cite{Prager2017,Lee2023} (see \methref{sec:Methods}). We emphasise that, of these parameters, we impose priors only on $L$, and nothing in our model \textit{a priori} favoured the results we find with respect to the other parameters, so the fact that we recover them is a good consistency check on our scenario. A further self-consistency check will become possible in upcoming years as more sensitive instruments such as the Cherenkov Telescope Array\cite{CTA19a} come online: these will be sensitive enough to image the IC emission from Terzan 5 at multiple energies. \autoref{fig:fpe_sample_solution} suggests that there should be clear variation in the morphology as a function of energy that should be detectable. Indeed, it is possible that such a signal could already be found in a detailed re-analysis of the existing HESS data, or by further HESS observations.

We can also verify that our scenario is consistent with the observed X-ray phenomenology of Terzan 5. \textit{Chandra} has detected diffuse X-ray emission from the position of the cluster, which may or may not be CRe-driven, with an estimated X-ray luminosity $L_X \approx 2\times 10^{33}$ erg s$^{-1}$ in the 0.7 - 10 keV band\cite{Eger10a}; no emission is detected from the location of the HESS emission peak, although this region is at the edge of the field of view and thus the sensitivity is lower than at the cluster position. Nonetheless, the non-detection is somewhat surprising\cite{Abramowski2011}, since the $p\sim 10$ TeV$/c$ CRe required to drive the emission detected by HESS should also produce synchrotron emission with a cutoff energy $E_\mathrm{sy} = 3 e B h p^2 \sqrt{1 - \mu_\mathrm{obs}^2} / 4 \pi m_e^3 c^3 = 0.66 \, (B/100\,\mu\mathrm{G}) (pc/10\,\mathrm{TeV})^2 \sqrt{1 - \mu_\mathrm{obs}^2}$ keV. However, it is difficult to translate the non-detection into a quantitative upper limit, because, in addition to reduced sensitivity due to the IC peak being at the edge of the \textit{Chandra} field of view, the line of sight to Terzan 5 is heavily absorbed. In \methref{sec:Methods} we estimate a hydrogen column of $N_\mathrm{H} = 2\times 10^{22}$ cm$^{-2}$ toward the cluster itself, corresponding to an optical depth of 3.5 at 1 keV, but the column varies with position, so the absorption optical depth to the region of IC emission is unknown.

We can nonetheless check whether our models are qualitatively consistent with the observed X-ray emission by computing the predicted X-ray spectrum from our best-fitting models, assuming the same absorption column as that measured toward Terzan 5 itself; see \methref{sec:Methods} for details. \edfref{fig:synchrotron} shows our prediction of the observable synchrotron emission. We find a huge range of possible X-ray luminosities: the 68\% confidence interval for the specific luminosity at 1 keV spans a factor of $\approx 1000$, including values significantly below the range observed by ref.~\cite{Eger10a}, and the 95\% interval extends to effectively zero luminosity. This broad range might at first seem surprising given the relatively small range of allowed IC luminosities (\autoref{fig:spectrum}), but can be understood as arising from the way that our uncertainties interact with the sharp energy dependence, roughly $\tau\propto E_\gamma^{-3.5}$, of the foreground opacity. Using our MCMC results to compute the value of $E_\mathrm{sy}$ at $p = p_\mathrm{eq}$ yields a 68\% confidence interval that goes from $\approx 1$ eV to $\approx 0.3$ keV (see \edtref{tab:posteriors}). For values near the upper limit we find levels of emission at energies $\gtrsim 1$ keV far too bright to be consistent with the non-detection of X-rays at the $\gamma$-ray maximum, but for cases closer to the low end of this range all of the synchrotron emission occurs at energies $\lesssim 0.1$ keV, where the optical depth is very large and thus none of the radiation reaches us. Given this huge range of possible outcomes, deeper X-ray searches toward Terzan 5 may not be able to detect anything. Conversely, however, given the sensitivity of the synchrotron emission to the values of $B$, $\mu_\mathrm{obs}$, and $p_\mathrm{eq}$, a detection would strongly constrain these parameters.

Our measured value of $K_\mu$ has important implications for CR transport. In particular, we find that the measured $K_\mu$ is consistent with the idea that it is the CR electrons themselves that excite the magnetic field perturbations on which they pitch-angle scatter via the gyro-resonant streaming instability\cite{Wentzel1968,Kulsrud1969,Kulsrud2005, Zweibel2017}. We show in the Supplementary Information that the streaming instability in the Terzan 5 magnetotail is likely in an unsaturated phase during which the CR pitch angle scattering rate should be comparable to the growth rate of Alfv\'en waves driven by CR streaming. For streaming speeds much larger than the Alfv\'en speed in the medium this is given by\cite{Kulsrud2005}
\begin{equation}
    \Gamma_\mathrm{si}(>p) \approx \Omega_B \frac{m_e}{m_p} \frac{n_\mathrm{CR}(>p)}{n_\mathrm{H}} \frac{v_\mathrm{str}(>p)}{v_\mathrm{A}},
    \label{eq:growth_rate}
\end{equation}
where $\Omega_B = eB/m_e c$ is the magnetic cyclotron frequency, $n_\mathrm{CR}(>p)$ and $v_\mathrm{str}(>p)$ are the number density and streaming speed of CRe with momentum $>p$, and $v_\mathrm{A} = B/\sqrt{4\pi n_\mathrm{H} \mu_\mathrm{H}}$ is the Alfv\'en speed, which we evaluate assuming the gas in the tail is fully ionised. In addition to the CR momentum $p$, this quantity depends on both the ambient gas density $n_\mathrm{H}$ and the cross-sectional area transverse to the field over which the CRs are spread, but for purposes of a rough estimate we define $\Gamma_0$ as the growth rate $\Gamma_\mathrm{si}$ evaluated at momentum $p_\mathrm{eq}$ (i.e., we use $n_\mathrm{CR}(>p_\mathrm{eq})$ and $v_\mathrm{str}(>p_\mathrm{eq})$) and assuming $n_\mathrm{H} = 1$ cm$^{-3}$ and that CRs are spread over a cylinder whose radius is given by the observed HESS source size in the transverse direction -- see \methref{sec:Methods} for details. We find $\Gamma_0 = 7.2_{-6.5}^{+56}\times 10^{-10}$ s$^{-1}$ (see \edtref{tab:posteriors}), so our confidence intervals for $K_\mu$ and $\Gamma_0$ overlap, consistent with the expectation in the self-confinement model that these quantities should be similar. \edfref{fig:corner} confirms that the joint posterior PDF of $\Gamma_0$ and $K_\mu$ has its maximum along the strip $\Gamma_0 \approx K_\mu$. We emphasise that nothing in the setup of our analysis requires this outcome, and the fact that we obtain it is strong circumstantial evidence that, at least in the Terzan 5 system, CRs are self-confined. By contrast, we also show in the Supplementary Information that confinement of CRs by extrinsic fast mode turbulence is unlikely to be able to explain the scattering rate we measure, although we cannot rule out the possibility that some other source of magnetic fluctuations, for example reconnection events, might produce a scattering rate that is also compatible with the one we measure. We also quantitatively compare the conditions in Terzan 5 to those in mean ISM conditions, and conclude that our finding suggests that self-confinement should also operate for ISM protons in warm ionised gas at energies of at least $\approx 35$ GeV. This conclusion is also consistent with the analysis of ref.~\cite{Thomas20a}, who find evidence for self-confinement of CR electrons, albeit in a very different energy range.

The method we have developed for measuring the pitch angle scattering rate, and thereby testing the CR self-confinement paradigm and other models for CR scattering and transport, is not limited to the case of Terzan 5. There are three other GCs -- Terzan 1, Terzan 2, and NGC 6838 -- that are both detected by \textit{Fermi} in the GeV band\cite{Song2021} and located $<500$ pc off the Galactic Plane\cite{Baumgardt21a}, where they might encounter ISM dense enough to produce MSP wind bow shocks $\lesssim 1$ pc from the cluster\cite{Bednarek2014}. An obvious step is to search for displaced TeV emission around these targets, although it is possible that a detection will require facilities more sensitive than HESS, since each of these GCs is roughly an order of magnitude dimmer than Terzan 5 at GeV energies. There are also sources associated with individual high velocity young or millisecond pulsars that exhibit displaced, non-thermal emission\cite{DeLuca2011,Pavan2014,Klingler2020,Crocker2022}. Of these, two seem particularly promising and direct analogues of the Terzan 5 system. First, the GeV-band $\gamma$-ray bright, but low spin-down power ($\sim 5.8 \times 10^{33}$ erg s$^{-1}$) and nearby ($\sim$500 pc) pulsar (PSR) J0357+3205 has an elongated ($\sim$1.3 pc) tail opposite its direction of proper motion\cite{Marelli2013}. The tail exhibits non-thermal X-ray emission whose surface brightness peaks discernibly ($\sim 4'\!\!.5$) away from the position of the pulsar\cite{DeLuca2011}. A synchrotron explanation for the observed emission requires  a high magnetic field amplitude ($\sim 50 \ \mu$G) and a population of radiating pairs with characteristic Lorentz factor $\gamma_e \sim 10^8$, both very similar to what we infer for Terzan 5. Second, the powerful ($\sim 1.8 \times 10^{36}$ erg s$^{-1}$) PSR J1809-1917\cite{Klingler2020}
exhibits an arcminute scale X-ray nebula extended both in the direction opposite to its motion and towards the associated TeV source HESS J1809-193\cite{Aharonian2023}, the centroid of which is offset from the pulsar itself\cite{Klingler2020}. We anticipate that more such sources are discoverable, in particular, by the LHAASO array\cite{Lhaaso2016} and, in the near future, by the Cherenkov Telescope Array\cite{CTA19a}. This raises the possibility of carrying out a much broader campaign of direct measurements of CR pitch angle scattering rates.

\begin{methods}
\label{sec:Methods}

To solve \autoref{eq:fpe} of the main text, we first non-dimensionalise it by defining $\tau = K_\mu t$, $\zeta = (K_\mu/c) z$, $\dot{\mathcal{N}} = \dot{N}/K_\mu$ and $q = p/p_\mathrm{eq}$, with $p_\mathrm{eq} = K_\mu t_{c,0} m_e c$, yielding the non-dimensional equation
\begin{eqnarray}
    \frac{\partial f}{\partial \tau} & = &
    -\mu \frac{\partial f}{\partial \zeta} + 
    \frac{\partial}{\partial \mu}\left[\left(1-\mu^2\right)\frac{\partial f}{\partial\mu}\right] +
    \frac{1}{q^2}\frac{\partial}{\partial q}\left[ \left(1-\mu^2\right)q^4 f\right] + {}
    \nonumber \\
    & & \dot{\mathcal{N}}\frac{dn}{dq} \delta(\zeta) \Theta(\mu-\mu_0).
    \label{eq:fpe_nondim}
\end{eqnarray}
We obtain a numerical solution to \autoref{eq:fpe_nondim} by noting that it is a Fokker-Planck equation (FPE), which has an equivalent It\^o stochastic differential equation\cite{Krumholz22a}
\begin{equation}
    d\chi_j = A_j(\boldsymbol{\chi}) + \left(\sqrt{K(\boldsymbol{\chi})}\, dW\right)_j,
    \label{eq:sde}
\end{equation}
where $\boldsymbol{\chi} = (\zeta,q,\mu)$ are the coordinates of a single probability packet in configuration space, the index $j = 1,2,3$ goes over the dimensions of this space, $d\mathbf{W}$ is the three-dimensional Wiener process, $\mathbf{A} = (\mu, q^2(1-\mu^2), -2\mu K_\mu)$ is the drift vector, $\mathbf{K} = (0, 0, 2K_\mu(1-\mu^2))$ is the vector diagonal of the diffusion tensor, and we apply reflecting boundary conditions at $\zeta = 0$. The ensemble solution to \autoref{eq:sde} for packets injected with an initial momentum distribution $dn/dq$ at a constant rate $\dot{\mathcal{N}}$ gives the solution to the corresponding FPE.

We can realise a significant computational gain by noting that \autoref{eq:sde} has a self-similarity property that removes the need to compute separate solutions for different initial momentum distributions $dn/dq$. The $\zeta$ and $\mu$ dimensions of this equation do not depend on $q$, and the $q$ dimension drifts as $A_q \propto q^2$, which means that a single packet trajectory $(\zeta(\tau), q(\tau), \mu(\tau))$ obtained by solving \autoref{eq:sde} can be rescaled to another valid solution with a different initial momentum $q'(\tau)$. In particular, one can readily verify by direct substitution that if some trajectory $q(\tau)$ satisfying $q(0) = 1$ and $q(\tau_1) = q_1$ is a solution to \autoref{eq:sde}, then the trajectory
\begin{equation}
    q'(\tau) = \frac{q_1 q'_1 \tau_1}{q_1 \tau_1 + q'_1(\tau-\tau_1)(1-q_1)}
\end{equation}
is also a valid solution (with $q'(\tau_1) = q'_1$) for any $q'_1$ satisfying $q_1' < q_1/(1-q_1)$. This relation allows us to transform a set of packets that solve \autoref{eq:sde} for $dn/dq = \delta(q-1)$, i.e., where all packets start with momentum $q = 1$, into a solution for an arbitrary choice of the initial momentum distribution. In particular, suppose that we find that some packet that started with momentum $q = 1$ at $\tau = 0$ has a momentum $q_1$ at some later time $\tau_1$. If we had instead drawn its initial momentum $q'_0$ from some distribution $dn/dq'_0$, then the resulting probability distribution of momenta $q'_1$ at time $\tau_1$ would be $dn/dq'_1 = (dq'_0/dq'_1)(dn/dq'_0) = (q'_0/q'_1)^2 (dn/dq'_0)$.

To generate solutions to \autoref{eq:fpe_nondim} for arbitrary momentum distributions, we therefore first use \textsc{criptic}\cite{Krumholz22a} to solve \autoref{eq:sde} for the case of CR sources that inject CRs at a rate of 1 per dimensionless time at position $\zeta = 0$ and momentum $q = 1$, with cosine pitch angles uniformly drawn from the interval $\mu = (\mu_0, 1)$; we use 16 sources whose values of $\mu_0$ range from $-0.5$ to 1 in steps of 0.1. We run this simulation in two stages: we first run to $\tau=100$ (long enough to reach steady state at all momenta $q\gtrsim 0.01$), injecting $10^6$ packets per dimensionless time $\tau$, and then from $\tau = 100$ to $110$ using $10^7$ packets per dimensionless time, yielding $2\times 10^8$ total packets by the end of the simulation; this approach of increasing $\Gamma$ with time ensures good sampling of both young, high momentum and old, low momentum packets. We then construct our solutions for the CR distribution function corresponding to each source at the final time using a kernel density estimate from the packet coordinates:
\begin{equation}
    f(\zeta,q,\mu) = \dot{\mathcal{N}} \sum_i w_i G(\zeta-\zeta_i, h_\zeta) G(\mu-\mu_i, h_\mu) \phi(q \mid q_i),
    \label{eq:fsol}
\end{equation}
where the sum runs over all the packets originating with a particular source, $w_i$ is the statistical weight in the \textsc{criptic} calculation (see ref.~\cite{Krumholz22a}), $G(x,h) = \exp(-x^2/h^2)/\sqrt{\pi h^2} $ is the Gaussian kernel with bandwidth $h$, and the distribution of present day momenta $q$ corresponding to a packet that has momentum $q_i$ in our simulation is
\begin{equation}
    \phi(q \mid q_i) = \left\{
    \begin{array}{ll}
    \left(q_0/q\right)^2 \left(dn/dq\right)_{q=q_0}, \quad & q < q_i/(1-q_i) \\
    0, & q > q_i/(1-q_i)
    \end{array}
    \right..
\end{equation}
Here $q_0 = (1 + q^{-1} - q_i^{-1})^{-1}$ is the initial momentum that the packet would have to have been assigned in order reach momentum $q$ at the end of the simulation, and $(dn/dq)_{q=q_0}$ is the distribution of injected momenta evaluated at $q_0$. For the purposes of numerical evaluation we adopt $h_\mu = 0.025$ in what follows. The corresponding dimensional distribution is given by an expression identical to \autoref{eq:fsol}, but with the pre-factor in front of the summation changed from $\dot{\mathcal{N}}$ to $\dot{N}/p_\mathrm{eq} c$. We can also generate solutions for values of $\mu_0$ that fall in between the points of our grid simply by linearly interpolating between the grid solutions, so for example we take $f_{\mu_0 = 0.78}(\zeta,q,\mu) = 0.2 f_{\mu_0 = 0.7}(\zeta,q,\mu) + 0.8 f_{\mu_0 = 0.8}(\zeta,q,\mu)$.

Given the distribution function, we can also compute a number of derived quantities that will be of interest in the remainder of our analysis. Two of these are the total line density (i.e., number of CRe per unit length) and streaming speed of CRe with momentum $>p$. If we consider some range $(z_0, z_1)$ along the field with $z_1 - z_0 \gg h_\zeta$ (which we can ensure by choosing sufficiently small $h_\zeta$), then we can compute the mean line density and streaming speed as
\begin{eqnarray}
    \left\langle\frac{dN(>p)}{dz}\right\rangle & = & \int_{z_0}^{z_1} \int_{p}^\infty \int_{-1}^1 f(z,p',\mu) \,d\mu \, dp' \, dz  
    \nonumber \\
    & \approx & \frac{\dot{N}}{c} \sum_i w_i \Phi(q \mid q_i) \\
    \left\langle v_\mathrm{str}(>p)\right\rangle & = & c \left\langle\frac{dN(>p)}{dz}\right\rangle^{-1} 
    \nonumber \\
    & & \qquad \int_{z_0}^{z_1} \int_{p}^\infty \int_{-1}^1 f(z,p',\mu) \mu \,d\mu \, dp' \, dz
    \nonumber \\
    & \approx & c \frac{\sum_i \mu_i w_i \Phi(q \mid q_i)}{\sum_i w_i \Phi(q \mid q_i)}
\end{eqnarray}
where the summations run over those packets satisfying $\zeta_0 < \zeta_i < \zeta_1$, and
\begin{equation}
    \Phi(q \mid q_i) \equiv \int_{q}^\infty \phi(q' \mid q_i) \, dq' =    
    \left\{
    \begin{array}{ll}
        \frac{\Gamma\left(k_p+1,\frac{q_{0}}{q_\mathrm{cut}}\right)}{\Gamma\left(k_p+1,\frac{q_\mathrm{min}}{q_\mathrm{cut}}\right)}, & q_i > q/(q + 1) \\
        0, & q_i \leq q/(q+1)
        \end{array},
    \right.
\end{equation}
where $\Gamma(a,x)$ is the upper incomplete $\Gamma$ function and $q_\mathrm{min} = p_\mathrm{min}/p_\mathrm{eq}$. The choice of $p_\mathrm{min}$ is relatively arbitrary, and has little effect on our results as long as it is smaller than the smallest energy for which we wish to compute emission; for the remainder of this work we fix $p_\mathrm{min} = 1$ GeV$/c$. For the purposes of estimating the growth rate of waves due to the streaming instability in \autoref{eq:growth_rate}, we compute rates for momenta $>p_\mathrm{eq}$, adopt $n_\mathrm{H} = 1$ cm$^{-3}$, $n_\mathrm{CR}(>p) = \langle dN(>p)/dz\rangle / \pi r_\mathrm{trans}^2$, where $r_\mathrm{trans} = 3.1$ pc is best-fit size of the minor axis of the HESS source\cite{HESS-Collaboration15a} at the $d_\mathrm{Ter5} = 6.62$ kpc cluster distance\cite{Baumgardt21a}, and use $\langle dN(>p)/dz\rangle$ and $\langle v_\mathrm{str}(>p)\rangle$ evaluated over the region from $z_0 = 0$ to $z_1 = d_\mathrm{Ter5} \theta_1 / \sqrt{1-\mu_\mathrm{obs}^2}$, where $\theta_1 = 4'\!\!.0$ is the location of the peak of the $\gamma$-ray emission observed by HESS\cite{Abramowski2011}; we define the growth rate evaluated under these assumptions as $\Gamma_0$. This quantity should provide a rough estimate of the rate of wave growth for the CRs for which loss and isotropisation are most closely matched, and in the regions over which the CRs are most strongly isotropising. The growth rate depends on the poorly-known density and transverse radius $r_\mathrm{trans}$ over which the CRs are distributed as $\Gamma_\mathrm{si} \propto n_\mathrm{H}^{-1/2} r_\mathrm{trans}^{-2}$.

The next step in our analysis is to compute the radiation produced by the CRe due to IC and synchrotron emission. Given the large CRe energies with which we are concerned, we adopt the ultrarelativistic limit whereby each CRe beams its radiation into an infinitesimally thin cone at angle $\theta = \cos^{-1}\mu$ relative to the magnetic field, and thus an observer whose line of sight makes an angle $\theta_\mathrm{obs}$ relative to the magnetic field (see \autoref{fig:Terzan5_schematic}) sees emission only from CRe with pitch angles $\mu = \mu_\mathrm{obs} = \cos\theta_\mathrm{obs}$. The time-averaged number of photons per unit photon energy $E_\gamma$ per unit time that will be received by an observer located within this cone due to synchrotron emission by a single CRe of momentum $q p_\mathrm{eq}$ is\cite{Rybicki86a}
\begin{equation}
    \frac{d\dot{N}_\mathrm{\gamma,sy}}{dE_\gamma} = e^{-\tau(E_\gamma)} \frac{\sqrt{3} e^3 B}{2 h p_\mathrm{eq} m_e c^3 \sqrt{1-\mu_\mathrm{obs}^2}} q_\gamma^{-1} F\left(\frac{q_\gamma}{q^2 q_\mathrm{sy}}\right),
\end{equation}
where $\tau(E_\gamma)$ is the optical depth of the foreground material to photons of energy $E_\gamma$ between the observer and the emitter, $F(x) = x \int_x^\infty K_{5/3}(\xi)\,d\xi$ is the synchrotron function, $K_{5/3}$ is the modified Bessel function of the second kind of order $5/3$, $q_\gamma = E_\gamma/p_\mathrm{eq} c$, and $q_\mathrm{sy} = 3 p_\mathrm{eq} e B h \sqrt{1-\mu_\mathrm{obs}^2}/4\pi m_e^3 c^4$ is the dimensionless synchrotron cutoff energy for CRs of momentum $p_\mathrm{eq}$. To compute $\tau(E_\gamma)$, must estimate the foreground hydrogen column $N_\mathrm{H}$. We adopt $N_\mathrm{H} = 2 \times 10^{22}$ cm$^{-2}$ based on the measured optical reddening towards Terzan 5 and the correlation between reddening and hydrogen column density\cite{Bahramian15a, Foight16a, Zhu17a}. We compute the corresponding value of $\tau(E_\gamma)$ using v2.3 of the TBabs model\cite{Wilms00a} as implemented in \textsc{Xspec} v12.13.1\cite{Arnaud96a}.

To obtain the equivalent expression for IC emission, we must first describe the radiation field with which the CRe interact, which we approximate as a sum of dilute blackbody components, each characterised by a dilution factor $W$ and colour temperature $T$. Five of these components are both isotropic and independent of position, and represent the cosmic microwave background and the background radiation field from Galactic dust and stars. The former has $W=1$ and $T=2.73$ K, while for the latter we adopt the four-component fit of ref.~\cite{Song2021}, which has $T=(40,200,800,4500)$ K and $W=(10^{-4}, 2\times 10^{-8}, 10^{-10}, 3\times 10^{-12})$; ref.~\cite{Song2021} in turn obtained these results from \textsc{galprop}\cite{Porter2017}. The sixth and final component is from the cluster itself, which has luminosity and half-light radius $L_\mathrm{c} = 3.75\times 10^5$ L$_\odot$ and $r_\mathrm{c} = 0.31$ pc \cite{Urquhart2020}; our luminosity estimate comes from the total mass of $9.4 \times 10^5$ M$_\odot$ and an assumed central value of the mass-to-light ratio M/L = $2.5 \pm 0.3 \mathrm{M}_\odot/\mathrm{L}_\odot$, taken from ref.~\cite{Baumgardt21a}. We therefore approximate the energy density of the cluster's radiation field experienced by CRe at position $z$ as $U_\mathrm{c} = L_\mathrm{c} / 4 \pi (z^2 + r_\mathrm{c}^2) c$, and the direction of the radiation field as being purely radially outward at all points; this description is clearly rough for $|z| \lesssim r_\mathrm{c}$, but this represents a small volume so the effect of the approximation is not large. Cluster light is dominated by relatively cool evolved stars, which we approximate with a blackbody spectrum with a colour temperature $T = 4500$ K, so the dilution factor of the cluster radiation field is $W = L_\mathrm{c} / 16 \pi \sigma_\mathrm{SB} T^4 (z^2 + r_\mathrm{c}^2)$, where $\sigma_\mathrm{SB}$ is the Stefan-Boltzmann constant. Given this description of the radiation field, we can calculate the rate of IC photon emission per unit time per unit photon energy per CRe as\cite{Krumholz22a, Khangulyan14a}
\begin{eqnarray}
    \frac{d\dot{N}_\mathrm{\gamma,IC}}{dE_\gamma} & = & \frac{\alpha^3}{8\pi\gamma_\mathrm{eq}^3}\left(\frac{1}{r_e p_\mathrm{eq}}\right) q^{-2} 
    \nonumber \\
    & & \quad \sum_i W_i \Upsilon_{\mathrm{eq},i}^2 F_{\mathrm{iso/mono}}(q_\gamma/q,\Upsilon_{\mathrm{eq},i},\mu_\gamma),
\end{eqnarray}
where the sum is over the components of the radiation field, $W_i$ and $T_i$ are the dilution factor and colour temperature of the $i$th component, $\alpha$ is the fine structure constant, $r_e$ is the classical electron radius, $\gamma_\mathrm{eq} = p_\mathrm{eq} / m_e c$ is the Lorentz factor of a CR of momentum $p_\mathrm{eq}$, $\Upsilon_{\mathrm{eq},i} = 4 p_\mathrm{eq} k_B T_i/m_e^2 c^3$ is the Klein-Nishina parameter for a CR of momentum $p_\mathrm{eq}$ scattering photons of energy $k_B T_i$, $F_\mathrm{iso/mono}$ is a dimensionless function describing the spectral shape for isotropic or mono-directional radiation fields, respectively, and for mono-directional radiation fields $\mu_\gamma = \mu_\mathrm{obs}$ is the cosine of the angle between the CR and initial photon directions. For isotropic radiation (all components of the radiation field except that of the cluster itself)
\begin{equation}
    F_\mathrm{iso}(y,\Upsilon_\mathrm{BB}) = \int_{y/[\Upsilon_\mathrm{BB}(1-y)]}^\infty \frac{x}{e^{x}-1} S(y,x,\Upsilon_\mathrm{BB})\, dx,
\end{equation}
where $S = 2\xi \ln \xi + (1 + 2\xi)(1-\xi) + (x\Upsilon_\mathrm{BB} \xi)^2/[2(1+x\Gamma_\mathrm{BB} \xi)]$ and $\xi = y/[x\Upsilon_\mathrm{BB}(1-y)]$, while for mono-directional radiation (the cluster field) we instead have 
\begin{eqnarray}
     \lefteqn{F_\mathrm{mono}(y,\Upsilon_\mathrm{BB},\mu_\gamma) = }
     \nonumber \\
     & & 
     \frac{y^2}{2(1-y)} \left[
    \mathrm{Li}_2(e^{-\xi_\mu}) - \xi_\mu \ln\left(1-e^{-\xi_\mu}\right)
    \right] + \mathrm{Li}_2(e^{-\xi_\mu}) + {}
    \nonumber \\ 
    & & \xi_\mu \ln(1 - e^{-\xi_\mu}) + 
    2 \xi_\mu^2 \int_{\xi_\mu}^\infty \frac{1}{x(e^x-1)} dx,
\end{eqnarray}
where $\xi_\mu = 2y/[\Upsilon_\mathrm{BB}(1-y)(1-\mu_\gamma)]$, and
$\mathrm{Li}_2$ is the dilogarithm function. 

Given $d\dot{N}_\gamma/dE_\gamma$, we can write the photon number flux per unit length along the magnetic field that reaches Earth as
\begin{eqnarray}
    \frac{d^2 \Phi_\gamma}{dE_\gamma\,dz} & = & \frac{1}{4\pi d_\mathrm{Ter5}^2}\int_0^\infty f(z,p,\mu_\mathrm{obs}) \frac{d\dot{N}_\gamma}{dE_\gamma} \, dp 
    \nonumber \\
   & = & \frac{\dot{N}}{4\pi d_\mathrm{Ter5}^2 c} \sum_i w_i G(\zeta-\zeta_i,h_\zeta) G(\mu_\mathrm{obs}-\mu_i,h_\mu) 
   \nonumber \\
   & & \quad \left\langle\frac{d\dot{N}_\gamma}{dE_\gamma}\right\rangle_{q_i}
\end{eqnarray}
where
\begin{equation}
    \left\langle\frac{d\dot{N}_\gamma}{dE_\gamma}\right\rangle_{q_i} = \int_{q_\gamma}^\infty \frac{d\dot{N}_\gamma}{dE_\gamma} \phi(q\mid q_i)\,dq.
\end{equation}
The quantity $\langle d\dot{N}_\gamma/dE_\gamma\rangle$ is the number of photons emitted per unit time per CRe averaged over the momentum distribution corresponding to a probability packet with momentum $q_i$ at the end of our simulation. While this quantity is the true emission, from the standpoint of comparing to the observed angular profile we must consider the instrumental point-spread function $P(\Delta\theta)$ and integrate the emission over some finite band in photon energy $(E_{\gamma,0},E_{\gamma,1})$. The band-integrated observed angular profile is then
\begin{equation}
    \frac{d\Phi_\gamma}{d\theta} = \frac{\dot{N}}{4\pi d_\mathrm{Ter5}^2 K_\mu} \sum_i w_i P(\theta - \theta_i) G(\mu_\mathrm{obs}-\mu_i, h_\mu) \left\langle \dot{N}_\gamma\right\rangle_{q_i}, 
    \label{eq:spec_angle}
\end{equation}
where $\langle \dot{N}_\gamma\rangle_{q_i} = \int_{E_{\gamma,0}}^{E_{\gamma,1}} \langle d\dot{N}_\gamma/dE_\gamma\rangle \, dE_\gamma$ is the photon emission rate integrated across the band, $\theta_i = \zeta_i (c/K_\mu d_\mathrm{Ter5}) \sqrt{1-\mu_\mathrm{obs}^2}$ is the angular position of the $i$th packet as seen from Earth, and this expression is valid in the limit of small $h_\zeta$. Following ref.~\cite{Aharonian06b}, we model the HESS PSF with the functional form
\begin{equation}
    P_\mathrm{HESS}(\Delta\theta) = N_\theta \left[\exp\left(-\frac{\Delta\theta}{2\sigma_1^2}\right) + A \exp\left(-\frac{\Delta\theta}{2\sigma_2^2}\right)\right],
\end{equation}
where $\sigma_1 = 2.76$ arcmin, $\sigma_2 = 7.2$ arcmin, $A=0.15$, and $N_\theta = 1 / [\sqrt{2\pi}(\sigma_1 + A \sigma_2)]$ is a normalisation factor chosen so that $\int P_\mathrm{HESS}(\Delta\theta)\,d\Delta\theta = 1$. Finally, we can further integrate over angle $\theta$ to obtain the total photon flux a given energy bin; since the PSF $P(\Delta\theta)$ is normalised to have unit integral, this is simply given by \autoref{eq:spec_angle} with the factor $P(\theta-\theta_i)$ removed.

Practical evaluation of the photon flux requires some care, because $\langle d\dot{N}/dE_\gamma\rangle_{q_i}$ and $\langle\dot{N}_\gamma\rangle_{q_i}$ are defined in terms of integrals that depend on the dimensionless momentum $q_i$ of the CR packet producing the emission (which determines the momentum distribution of the CRs contributing to the emission), the photon energy or energy range being observed, the parameters $k_p$ and $p_\mathrm{cut}$ that describe the injection momentum distribution, the momentum scale factor $p_\mathrm{eq}$, the cosine of the photon-CR angle $\mu_\gamma$ (for IC scattering by mono-directional radiation fields), and the magnetic field strength $B$ (for synchrotron emission). To avoid having to evaluate this integral repeatedly as we move through parameter space while attempting to fit the observations, we build an interpolation table. We use the \textsc{cubature} package\cite{cubature} to evaluate $\langle d\dot{N}_\gamma/dE_\gamma\rangle_{q_i}$ and $\langle\dot{N}_\gamma\rangle_{q_i}$ over a grid of values in $k_p$ from $-3$ to 0 in steps of 0.1, in $\log (p_\mathrm{cut} c / \mathrm{GeV})$ from 2 to 7 in steps of 0.1, in $\log (p_\mathrm{eq} c/\mathrm{GeV})$ from 1 to 6 in steps of 0.1, in $\mu_\gamma$ from $-0.95$ to 0.95 in steps of 0.1, in $\log (B/\mu\mathrm{G})$ from 1 to 3 in steps of 0.1, and in $\log q_i$ from $-2.1$ (just below the lowest momentum produced in our simulations) to 0 in steps of 0.025 dex. We then linearly interpolate on the resulting table to compute the emission for any CR momentum $q_i$ and any combination of parameters $(\mu_0, k_p, p_\mathrm{cut}, p_\mathrm{eq}, K_\mu, \dot{N}, \mu_\mathrm{obs})$. Note that these parameters are degenerate with, and uniquely determine, the magnetic field strength and the total CR luminosity:
\begin{eqnarray}
    B & = & \sqrt{\frac{4\pi^2 m_e^2 c^2 K_\mu}{p_\mathrm{eq} \sigma_\mathrm{T}}} \\
    L & = & \dot{N} p_\mathrm{eq} c q_\mathrm{cut} \frac{\Gamma(k_p+2,q_{\mathrm{min}}/q_\mathrm{cut})}{\Gamma(k_p+1,q_{\mathrm{min}}/q_\mathrm{cut})}.
\end{eqnarray}

The final step in our analysis is to determine what range of parameter values is consistent with the observations. To this end we define a goodness-of-fit statistic that combines the spectral and angular constraints on the IC emission. The spectral constraints are relatively straightforward: as discussed in the main text, the low-energy emission is likely dominated by MSP magnetospheric signal, so we fit the predicted IC to emission to the spectrum observed by \textit{Fermi} plus HESS at energies $E_\gamma > 18.5$ GeV, which we obtain from ref.~\cite{Song2021}. The authors of this work used the spectral data obtained by the HESS Collaboration \cite{Abramowski2011} in the TeV band and obtained their own Terzan 5 spectrum using 8 years of Pass 8 {\it Fermi}-LAT data accumulated from 2008 August 4 to 2016 August 2 (the same data as used in the 4FGL catalogue); the reader is referred to 
further details of the ref.~\cite{Song2021} for further details of the {\it Fermi}-LAT data analysis. The angular constraints are somewhat more challenging; ideally we would compare the forward-modelled angular profile $d\Phi_\gamma/d\theta$ to the observations directly, but ref.~\cite{Abramowski2011} does not provide tabulated angular profiles, only the result of a Gaussian fit to the integrated-light angular profile at energies $>1$ TeV. This fit provides the centre and dispersion of the angular distribution, $(\theta_1, \theta_2) = (4'\!\!.0, 9'\!\!.6)$, and corresponding uncertainties $(\Delta\theta_1,\Delta\theta_2) = (28''\!\!\!.5, 36'')$. (Note that the uncertainties reported in ref. \cite{Abramowski2011} contain a typographical error; the values we use have been corrected for this.) We therefore take our goodness of fit statistic to be
\begin{eqnarray}
    \chi^2 & = & \sum_i \left(\frac{\Phi_{\gamma,i} - \Phi_{\gamma,i,\mathrm{mod}}}{\Delta \Phi_{\gamma,i}}\right)^2 + \left(\frac{\theta_1 - \theta_{1,\mathrm{mod}}}{\Delta\theta_1}\right)^2 + {}
    \nonumber \\
    & &  \left(\frac{\theta_2 - \theta_{2,\mathrm{mod}}}{\Delta\theta_2}\right)^2,
\end{eqnarray}
where $\Phi_{i,\mathrm{mod}}$ is the model-predicted photon flux in band $i$, $\theta_{1,\mathrm{mod}}$ is the model-predicted location of maximum emission (i.e., the maximum of $d\Phi_\gamma/d\theta$), and $\theta_{2,\mathrm{mod}}$ is the second moment of the angular distribution about this maximum,
\begin{equation}
    \theta_{2,\mathrm{mod}} = \sqrt{\frac{\sum_i W_i \left[(\theta_i - \theta_{1,\mathrm{mod}})^2 (\sigma_1 + A\sigma_2) + \sigma_1^3 + A \sigma_2^3\right] }{\sum_i W_i (\sigma_1 + A\sigma_2)}}.
\end{equation}
Here $W_i \equiv w_i G(\mu_\mathrm{obs}-\mu_i,h_\mu) \langle \dot{N}_\gamma\rangle_{q_i}$ is the weight of the $i$th packet, and $\langle \dot{N}_\gamma\rangle_{q_i}$ evaluated for $E_\gamma > 1$ TeV is the photon number flux per CR integrated over the $>1$ TeV band. We define a corresponding log-likelihood $\ln \mathcal{L} = -\chi^2/2$. For the purposes of assessing goodness of fit, we also define a reduced $\chi^2$, defined as $\chi^2_\mathrm{red} = \chi^2/(N_\mathrm{obs} - N_\mathrm{dof})$, where we have $N_\mathrm{obs} = 13$ observations (11 measured photon fluxes $\Phi_{\gamma,i}$ plus $\theta_1$ and $\theta_2$) and $N_\mathrm{dof} = 7$ model degrees of freedom ($K_\mu$, $\mu_\mathrm{obs}$, $p_\mathrm{eq}$, $p_\mathrm{cut}$, $k_p$, $\mu_\mathrm{inj}$, and $\dot{N}$).

From the likelihood we determine the posterior probability distribution function of the parameters $(\mu_0, k_p, p_\mathrm{cut}, p_\mathrm{eq}, K_\mu, \mu_\mathrm{obs})$ from a Markov Chain Monte Carlo (MCMC) calculation using the package \textsc{emcee}\cite{Foreman-Mackey13a}; note that in order to reduce computation time we do not include the total CR injection rate $\dot{N}$ in our MCMC analysis, because it represents an overall normalisation and thus its maximum likelihood and posterior distribution can be computed analytically. We evaluate the likelihood at any point in our reduced parameter space $(\mu_0, k_p, p_\mathrm{cut}, p_\mathrm{eq}, K_\mu, \mu_\mathrm{obs})$ using the maximum-likelihood value of $\dot{N}$. For the purposes of our MCMC, we adopt flat priors over all $\log K_\mu$, and for the remaining parameters we adopt flat priors over the range in parameter space covered by our interpolation grid: $k_p$ from $-3$ to 0, $\log [p_\mathrm{cut}/(\mathrm{GeV}/c)]$ from 2 to 7, $\log [p_\mathrm{eq}/(\mathrm{GeV}/c)]$ from 1 to 6, $\mu_\mathrm{obs}$ from $-0.95$ to $0.95$, and $\mu_0$ from $-0.5$ to 1. We also impose as an additional prior constraint that $L<10^{38}$ erg s$^{-1}$, which is a generous upper limit on the total energy budget of the aggregated pulsar winds. The latter can be estimated on the basis of two, independent considerations. First, on the basis of the measured\cite{Song2021} $\sim 10^{35}$ erg s$^{-1}$ luminosity of the GeV band signal (attributed to the aggregated magnetospheric, curvature radiation from Terzan 5's MSP population) we expect\cite{Abdo2013,Bednarek2014,Crocker2022} a wind power in the range $10^{36}$--$10^{37}$ erg s$^{-1}$. Second, we can use the period and period derivatives for MSPs in Terzan 5 that have been directly measured at radio frequencies\cite{Prager2017,Lee2023} to estimate an aggregate spin-down power of $\sim 10^{37}$ erg s$^{-1}$. Given the effect of the cluster potential on the {\it measured} period derivatives of Terzan 5 MSPs, this is only a rough estimate and necessarily a lower limit because it does not attempt to correct for individual MSPs that do not have radio measurements, and assigns zero luminosity to MSPs whose measured period derivative is positive. Adopting a prior upper limit on the luminosity is important because for $k_p > -2$, which as discussed in the main text is expected based on the shape of the observed photon spectrum, most of the radiatied power lies near $p_\mathrm{cut}$. However, CRs injected with $p \gg p_\mathrm{eq}$ generally produce no observable emission, because they cool before they have time to be scattered into our line of sight. Consequently the $\gamma$-ray data permit $L$ and $p_\mathrm{cut}$ to be arbitrarily large, so long as $p_\mathrm{cut} \gg p_\mathrm{eq}$ and thus none of the ultra-high energy CRe produce visible emission. Imposing a prior on $L$ prevents the MCMC from wandering into this unphysical scenario.

We run the MCMC using 104 walkers for $\approx 17,000$ iterations. We estimate the autocorrelation time of the chains for each quantity by fitting an autoregressive model following the approach recommended in the \textsc{emcee} documentation at \url{https://emcee.readthedocs.io/en/stable/tutorials/autocorr/}. We find autocorrelation times ranging from $\approx 420$ steps (for $p_\mathrm{cut}$) to $\approx 770$ steps (for $\mu_\mathrm{obs}$). We therefore take the autocorrelation time to be 800 iterations for safety, meaning that our run covers more than 20 autocorrelation times. We discard the first five autocorrelation times for burn-in and derive our posterior PDFs from the remaining samples.

\end{methods}

\begin{si}

In this supplementary information we provide more detailed calculations for some points made in the main text. In \sisecref{sec:bowshock} we estimate the luminosity of the acceleration region near the bow shock of Terzan 5, in \sisecref{sec:alternatives} we consider alternative scenarios for explaining the displacement of the TeV signal, and in \sisecref{sec:transport} we work through further details of the streaming instability and of alternative CR scattering mechanisms, and use this analysis explore the implications of our results for CR transport in other environments.

\section{Emission from the bow shock acceleration region\\}
\label{sec:bowshock}

In our calculation we have assumed the observable emission is mostly from the magnetotail where the accelerated, pitch angle-anisotropic CR population isotropises. However, in this scenario there is also a population of CRe in the acceleration region near the bow shock, and this population has a much more isotropic pitch angle distribution. Thus in principle they should also produce inverse Compton emission, and if this emission is significant then it should be included in our model. However, we can show that this acceleration region emission should be significantly dimmer than the magnetotail emission, justifying our focus on the tail, through a fairly simple calculation.

Suppose that the steady-state total energy of CRe in the acceleration region is $E_{e}$, and let $t_\mathrm{IC}$ and $t_\mathrm{esc}$ be the inverse Compton cooling time and escape time in this region, respectively. In this case the expected inverse Compton luminosity from the acceleration region is
\begin{equation}
    L_\mathrm{acc} \approx \frac{E_{e}}{t_\mathrm{IC}} \approx \frac{t_\mathrm{esc}}{t_\mathrm{IC}} L_\mathrm{CR},
    \label{eq:Lacc}
\end{equation}
where $L_\mathrm{CR}$ is the energy per unit time injected into CRe by the acceleration process. For the $\gtrsim 10$ TeV CRe capable of producing the $\gtrsim 1$ TeV photons whose displacement has been observed by HESS, using the model of the radiation field described in Methods, and for an accelerator region located $\approx 1$ pc from the cluster centre, the dominant radiation field for IC losses is the cosmic microwave background (CMB) rather than the cluster radiation field -- the cluster field has a higher energy density, but losses caused by it are strongly suppressed by Klein-Nishina corrections. Consequently, we adopt
\begin{equation}
    t_\mathrm{IC} = \frac{3 m_e c}{4 \gamma \sigma_T a_R T_\mathrm{CMB}^4},
\end{equation}
where $T_\mathrm{CMB}=2.73$ K is the CMB temperature and $\gamma$ is the CRe Lorentz factor.

Following ref.~\cite{Bykov2017}, we estimate the escape time as
\begin{equation}
    t_\mathrm{esc} \approx \frac{R_\mathrm{acc}^2}{\eta K_\mathrm{Bohm}} = \frac{3 e B R_\mathrm{acc}^2}{\eta \gamma m_e c^3}, 
\end{equation}
where $R_\mathrm{acc}$ is the characteristic size of the acceleration region bounded on one side by the bow shock and on the other by the pulsar wind termination shock, $K_\mathrm{Bohm} \equiv r_\mathrm{g} c / 3$ is the Bohm diffusion rate for a CRe with gyroradius $r_\mathrm{g} = \gamma m_e c^2/eB$, $B$ is the magnetic field strength, and $\eta$ is a numerical factor $\geq 1$. To estimate the acceleration region size, we note that we expect the cutoff momentum (which is returned by our fit to the $\gamma$-ray spectrum) to be set by the Hillas Criterion, which implies that accelerators cannot drive particles to momenta so large that the gyroradius exceeds the geometric size of the accelerator; we express this condition as
\begin{equation}
    R_\mathrm{acc} = \xi \frac{\gamma_\mathrm{cut} m_e c^2}{e B} = 1.1\times 10^{-3} p_\mathrm{cut} B_2^{-1} \xi \mbox{ pc},
    \label{eq:racc}
\end{equation}
where $\gamma_\mathrm{cut}$ is the Lorentz factor of a CR with momentum $p_\mathrm{cut}$, $p_\mathrm{cut,5} = p_\mathrm{cut}/10^5$ GeV$/c$, $B_2 = B / 100$ $\mu$G, and $\xi$ is a numerical factor, with $\xi = 1$ corresponding to setting the cutoff momentum to exactly the Hillas Criterion value. The values of $p_\mathrm{cut}$ and $B$ to which we have scaled the numerical results are close to the 50th percentile values returned by our fits to the $\gamma$-ray data. We can also place an upper limit on the possible value of $\xi$ by noting that no accelerated particles will escape the acceleration region at all unless the escape time is significantly smaller than the synchrotron loss time, $t_\mathrm{sync} = 6\pi m_e c / \gamma \sigma_T B^2$. This condition implies
\begin{equation}
    \xi \ll \sqrt{\frac{2\pi e \eta}{\sigma_T B \gamma_\mathrm{cut}^2}} \approx 34 \sqrt{\eta} B_2^{-1/2} p_\mathrm{cut,5}^{-1},
    \label{eq:xi_lim}
\end{equation}
Thus we expect that $\xi$ cannot be too far from unity.

Inserting our values for $t_\mathrm{IC}$, $t_\mathrm{esc}$, and $R_\mathrm{acc}$ into \autoref{eq:Lacc}, we obtain
\begin{equation}
    L_\mathrm{acc} \approx \frac{4 \sigma_T a_R T_\mathrm{CMB}^4 \gamma_\mathrm{cut}^2 \xi^2}{e B \eta} L_\mathrm{CR} = 8.9\times 10^{-7} p_\mathrm{cut,5}^2 B_2^{-1} \frac{\xi^2}{\eta} L_\mathrm{CR}.
\end{equation}
The corresponding energy-integrated flux received at the Sun from the accelerator region is
\begin{equation}
    F_\mathrm{acc} = \frac{L_\mathrm{acc}}{4\pi d_\mathrm{Ter5}^2} = 3.4\times 10^{-12} p_\mathrm{cut,5}^2 B_2^{-1} L_\mathrm{CR,37.5} \frac{\xi^2}{\eta}\mbox{ GeV cm}^{-2}\mbox{ s}^{-1},
    \label{eq:Facc_central}
\end{equation}
where $L_\mathrm{CR,37.5} = L_\mathrm{CR} / 10^{37.5}$ erg s$^{-1}$, again close to the 50th percentile value derived from our fit to the $\gamma$-ray data. Our limit from \autoref{eq:xi_lim} further implies that
\begin{equation}
    F_\mathrm{acc} \ll 4.0 \times 10^{-9} B_2^{-2} L_\mathrm{CR,37.5}\mbox{ GeV cm}^{-2}\mbox{ s}^{-1}.
    \label{eq:Facc_lim}
\end{equation}
Comparing to Figure 1 of the main paper, we see that our central estimate, \autoref{eq:Facc_central}, implies a $\gamma$-ray flux from the accelerator region that is a factor of $\sim 100$ smaller than the measured energy-integrated flux of the displaced $\gamma$-ray emission observed by HESS in the TeV band, and our upper limit from \autoref{eq:Facc_lim} implies that even if we choose the most extreme parameters possible to maximise the luminosity of the accelerator, we still end up with an accelerator luminosity smaller than the observed HESS signal. We therefore conclude that we are justified in neglecting emission from the accelerator region when modelling the HESS signal from Terzan 5.

\section{Alternative scenarios\\}
\label{sec:alternatives}

Here we consider two alternative scenarios to explain the displaced TeV signal, and discuss our reasons for thinking each of them to be less likely than the scenario proposed in the main text.\\

\subsection{Curvature in the magnetotail\\}
\label{ssec:curvature}

If the cosmic ray population in the magnetotail were anisotropic but did not undergo any pitch angle scattering at all, one could still produce displaced TeV emission if the magnetotail itself were not straight. In this scenario, the IC radiation beam is always aligned with the tail, and close to the cluster the tail and the radiation beam point away from Earth. However, as one proceeds down the tail it kinks or curves so that it points toward us, allowing us to see the radiation beam from the CRe.

The challenge in this scenario is to explain why the tail should contain a bend. It is important to keep in mind that the cluster is moving through the ISM far faster than any plausible estimate of the local gas velocity dispersion, so to first order the ISM is a static background and the cluster and its wind is a cannonball passing through it, carving a cavity that is bounded by swept-up gas and magnetic fields as it does so; the magnetotail consists of the magnetic fields on the wall of this cavity.

We consider three distinct possible causes for the tail to bend. The first and most easily dismissed is that the cluster itself has taken a curved trajectory through the ISM due to gravitational acceleration by the Milky Way. To see why this does not work, note that the displaced $\gamma$-ray signal is $d \approx 10$ pc from the cluster, and if Terzan 5 is moving at a speed $v_\mathrm{T5}\sim 100$ km s$^{-1}$ relative to the ISM it has covered this distance in $d/v_\mathrm{T5} \sim 0.1$ Myr. By contrast, the orbital period is $20-302$ Myr\cite{Baumgardt21a}. Thus over the time from when Terzan 5 passed the location of the TeV $\gamma$-ray signal to now, it has completed only $\lesssim 1\%$ of its orbit. Thus the curvature of its path is negligible.

A second possible cause of bending that we can dismiss almost as quickly is due to forcing by motions in the Milky Way's ISM. To evaluate this possibility, consider a reference frame centred on and comoving with Terzan 5, and with the velocity of the interstellar material that Terzan 5 is encountering now pointing in the $+x$ direction, so that the magnetotail extends from $x\approx 0$ along the positive $x$ axis. If the background ISM did not have a uniform velocity along the extent of the tail, then in our reference frame the external ISM could have a velocity $v_\mathrm{tr}$ transverse to the magnetotail, which would exert a ram pressure on it, potentially causing it to bend. However, the resulting deflection cannot be significant. To see this, first note that over the distance $d \approx 10$ pc of interest, the observed velocity dispersion of the Milky Way's ISM is $\sim 3$ km s$^{-1}$ (equation 5 of ref.~\cite{Kalberla09a}), so $v_\mathrm{tr}\lesssim 3$ km s$^{-1}$. Now consider fluid that enters the magnetotail, traveling with velocity $v_\mathrm{tail} \sim v_\mathrm{T5} \sim 100$ km s$^{-1}$ in the cluster rest frame. This fluid element requires a time $t = d/v_\mathrm{tail}$ to travel down the tail, and even if it has zero inertia and is instantaneously accelerated to $v_\mathrm{tr}$ in the transverse direction, the total transverse displacement it will suffer is $v_\mathrm{tr} t$, and thus the ratio of the transverse displacement to the distance down the tail is $v_\mathrm{tr} / v_\mathrm{tail} \sim 0.03$; this corresponds to a bend of only $2^\circ$.

The final possibility we consider is that the tail could be bent by some internal instability, independent of external forcing; the Kelvin-Helmholz instabilty seems a natural candidate, and large-scale reconnection events is another possibility. However, simulations of pulsar bow shock magnetotails extending out to many tens of standoff radii (well past our $d\approx 10$ pc scale of interest) show no evidence for large-scale bending due to KH, reconnection, or any other instability, even in cases when the post-shock region becomes turbulent or where the pulsar wind is anisotropic\cite{Barkov19a, Olmi19a}; instabilities certainly appear in some cases, but they appear to drive small-scale turbulence, not large-scale kinks. The absence of KH-induced bending of the magnetotail in particular can be understood in the light of the extensive analysis of the supersonic KH instability in cylindrical structures that exists in the literature\cite{Mandelker16a, Padnos18a}. A key result from this work is that growth of the KH instability is strongly suppressed when the velocity difference across the interface is supersonic, as is certainly the case here, because the ultimate driver of the instability is that ripples at interfaces cause pressure perturbations -- overpressure upstream of wave crests, underpressure downstream -- and these pressure perturbations in turn cause the surface ripples to grow. This mechanism is suppressed for supersonic flows because a fluid element that is advecting supersonically passes many wave crests and troughs per sound crossing time of the perturbation, causing the positive and negative pressure perturbations to average to zero.

We therefore conclude that bends in the magnetotail are unlikely to explain the displaced TeV $\gamma$-ray signal from Terzan 5 because there is no good candidate mechanism to induce such bends. However, a definite test of this possibility can be made with future $\gamma$-ray data, either from further HESS observations or from CTA: in the bending scenario, the displacement and morphology of the IC emission should be independent of $\gamma$-ray energy, since a bend in the magnetic field will deflect particles of all energies equally, while in our preferred scenario the morphology will depend on energy.\\

\subsection{Particle acceleration by reconnection\\}

A second alternative scenario is that CRe are injected into the magnetotail with an isotropic pitch angle distribution, but that we initially do not see them in the TeV band accessible to HESS because they are insufficiently energetic. However, at some point down the magnetotail there is a reconnection event, which accelerates the CRe up to TeV energies and creates a ``hot spot'' that is visible to HESS and is displaced away from the cluster.

Relativistic reconnection is certainly capable of accelerating particles to the $\gtrsim 10$ TeV energies required to explain the observed emission, and for sufficiently magnetised flows can also produce the very hard particle spectra required by the observed photon spectrum\cite{Sironi14a, Kagan15a}. However, this scenario immediately encounters an observational problem: if there is a population of CRe along the full length of the magnetotail with an isotropic pitch angle distribution, then even if this population does not produce IC emission at the TeV energies detectable by HESS until it is re-energized by reconnection, it should still produce GeV emission visible to \textit{Fermi-LAT}. However, no such emission is detected. Quantitatively, ref.~\cite{Ndiyavala2019} re-analyse the \textit{Fermi-LAT} pass 8 data and find a point source with a centroid $0.\!\!'66$ from the cluster's optical centre and an estimated size of $\approx 1.\!\!'6$, with no evidence in the residuals for any additional sources. The position is consistent (given \textit{Fermi}'s angular resolution) with emission coincident with the cluster centre, and is clearly inconsistent with the centroid of the HESS signal. By contrast, if re-acceleration were the primary source of the TeV CRs, we would expect this process to produce a CR index $k_p$ no harder than $k_p = -2$. The resulting photon emission should have a spectral index comparable to or softer than this (softer due to Klein-Nishina suppression becoming stronger at higher energies), and therefore the spectrum $E_\gamma^2 \, dN_\gamma/dE_\gamma$ should be falling from GeV to TeV energies, implying the presence of a GeV source at least as bright as the HESS TeV source, and thus no more than an order of magnitude fainter than the curvature radiation at GeV energies seen from Terzan 5 itself (cf.~Figure 1 of the main text). Since the curvature radiation at the cluster position is detected at $\approx 70\sigma$ significance by \textit{Fermi}\cite{Abdollahi22a}, if there were GeV emission this bright at the location of the HESS source it should have been detected at $\gtrsim 7\sigma$ significance. The non-detection by ref.~\cite{Ndiyavala2019} is therefore hard to reconcile with a reacceleration scenario.

A second potential challenge to this scenario, but one that is hard to evaluate quantitatively, arises from the X-ray non-detection at the position of the TeV emission. The challenge is that, if $\sim 10^{34}-10^{35}$ erg s$^{-1}$ of energy is being deposited in non-thermal particles $\sim 10$ pc down the magnetotail, presumably there should be a similar amount of energy being deposited in thermal plasma, which will then radiate in X-rays. Given the very strong foreground extinction, whether this emission would be detectable or not depends sensitively on the plasma temperature distribution -- if the plasma is relatively cool, $k_B T \lesssim 1$ keV, it will be hidden by extinction, but if any significant component of it is hotter and radiates at $\gtrsim 1$ keV energies, this would be easily detectable (see \edfref{fig:synchrotron}). At present, relativistic reconnection models are not able to make quantitative predictions for the temperature distribution of the thermal plasma, so it is difficult to evaluate quantitatively whether this is a serious problem for a reconnection model, but it is a concern.

\section{Detailed analysis of cosmic ray scattering mechanisms, and implications for transport in other environments\\}
\label{sec:transport}

Here we provide a more detailed analysis of some aspects of CR scattering and transport. We first verify that the Terzan 5 magetotail should be in the growing rather than the saturated limit of the streaming instability, and check whether the scattering rate we have measured could be due to extrinsic turbulence instead. We then evaluate the implications of our findings for CR transport in two other environments: the warm ionised ISM of the Galaxy and the environments around other TeV sources, the so-called TeV halos.\\

\subsection{Wave damping and the streaming instability\\}
\label{ssec:saturated}

In the main manuscript we treat the streaming instability in Terzan 5 as being in the initial growth phase. Here we explain our reasoning for this conclusion. We begin by estimating the damping rate for waves that are resonant with CRe of momentum $p_\mathrm{eq}$. The Terzan 5 magntotail almost certainly has plasma $\beta \ll 1$, and in such gas the dominant damping mechanism is likely to be turbulent damping, which produces a damping rate (ref.~\cite{Lazarian16a}; however, see also ref.~\cite{Cerri24a}, who predict a smaller damping rate)
\begin{equation}
    \Gamma_\mathrm{turb} \approx \mathcal{M}_A^2\frac{v_A}{\sqrt{r_\mathrm{g} r_\mathrm{out}}}
    \label{eq:gamma_turb}
\end{equation}
for CRs of momentum $p$, where $\mathcal{M}_A = \sigma_\mathrm{v}/v_A$ and $r_\mathrm{out}$ are the Alfv\'en Mach number and outer scale of the turbulence, $r_\mathrm{g} = pc/eB$ is the CR gyroradius, $\sigma_\mathrm{v}$ is the gas velocity dispersion, and $v_A = B/\sqrt{4\pi \mu_\mathrm{H} n_\mathrm{H}}$ is the Alfv\'en speed. If we evaluate this using $r_\mathrm{out} = r_\mathrm{trans} = 3.1$ pc, adopt a velocity dispersion equal to the thermal gas sound speed at $T=10^4$ K, $\sigma_v = \sqrt{k_B T / \mu m_\mathrm{H}} \approx 11$ km s$^{-1}$ assuming $\mu = 0.64$ for ionised ISM and for our assumed $n_\mathrm{H} \sim 1$ cm$^{-3}$ and our 50th percentile values of $p_\mathrm{eq}$ and $B$, we have $\mathcal{M}_A \approx 0.06$, and we find $\Gamma_\mathrm{turb}\approx 2\times 10^{-12}$ s$^{-1}$, roughly 1.5 dex smaller than $\Gamma_\mathrm{si} \sim 10^{-10}$ s$^{-1}$ (which is also close to the synchrotron loss rate $\Gamma_\mathrm{sy}$ at $p = p_\mathrm{eq}$). Thus streaming instability can grow in the magnetotail.

The remaining question is whether the instability will be saturated, in which case the CR propagation speed should be reduced to $v_A$ and the population fully isotropised, or still growing, in which case our estimate of $\Gamma_\mathrm{si}$ from the main text is reasonable. The growth timescale we have computed is $\Gamma_\mathrm{si} \sim \mathrm{kyr}$, which is substantially smaller than the $\sim 100$ kyr that it takes Alfv\'en waves traveling at $v_A$ to traverse the magnetotail, so if the illumination of the magnetotail by the accelerator is steady then we should be in the saturated limit. However, there is a problem with such a scenario: if the CRe are slowed to velocity $v_A$, then over a synchrotron loss time they will travel only $v_A / \Gamma_\mathrm{sy}\sim 0.05$ pc, smaller than the size of the cluster and approaching the size of the accelerator (c.f.~\autoref{eq:racc}). Thus in any any part of the magnetotail where the Alfv\'en waves are at saturated levels there will be no CRe entering the tail at all; CRe can enter the magnetotail in the first place only in locations where the Alv\'en waves are far from saturated. Moreover, we expect such regions of relatively ``fresh'' magnetic field to exist in the magnetotail. The accelerator is likely to fluctuate in the location and rate of CRe production on scales comparable to its crossing time; since the cluster is sweeping up ISM at $\sim 100$ km s$^{-1}$, this time is only $\sim 100$ yr for the $\sim 0.01$ pc accelerator region and $\sim 1$ kyr for the $\sim 0.1$ pc cluster core. Thus we expect fluctuations in particle acceleration on timescales comparable to the growth timescale, suggesting a picture in which any region that is continuously illuminated by the accelerator becomes ``clogged'' with saturated Alfv\'en waves and does not contribute to the magnetotail, but at the same time new regions of unperturbed magnetic field are constantly being illuminated and allowing particles to escape. Clogged regions of saturated Alfv\'en waves also unclog on timescale as $\sim 30$ kyr as the waves within them damp. This leads us to a picture where entry of particles into the magnetotail is primarily via regions where Alfv\'en waves have not yet reached saturation.\\

\subsection{Are the Terzan 5 results compatible with scattering by extrinsic turbulence?\\}
\label{ssec:extrinsic}

We show in the main text that scattering by streaming instability-induced turbulence can naturally explain the measured value of the diffusion coefficient in Terzan 5, but it is interesting to consider whether this value might also be explained by scattering off waves that are part of a turbulent cascade from larger scales. As pointed out by ref.~\cite{Farmer04a}, the anisotropy of magnetohydrodynamic turbulence on small scales strongly inhibits the ability of Alfv\'enic fluctuations to scatter CRs. Consequently, scattering is likely dominated by either by gyroresonant scattering with fast modes or by transit-time damping\cite{Yan04a,Yan08a,Kempski22a}. For ultrarelativistic particles, the gyroresonant scattering rate can be written\cite{Yan08a}
\begin{eqnarray}
    K_\mu^\mathrm{G} & = & \frac{c}{r_\mathrm{out}} \frac{\sqrt{\pi}(1-\mu^2) \mathcal{M}_A^2}{2 R^2} \int_0^1 d\xi \int_1^{k_\mathrm{max} r_\mathrm{out}} dx\,
    \nonumber \\
    & & 
     \frac{\xi}{x^{5/2}\Delta \mu}\left[J_1'(w)\right]^2 \exp\left[-\frac{\left(\mu - 1/x\xi R\right)^2}{\Delta\mu^2}\right],
    \label{eq:KmuG}
\end{eqnarray}
where $\mathcal{M}_A$ is the Alfv\'en Mach number of the fast mode turbulence on the outer scale $r_\mathrm{out}$, $\mu$ is the cosine of the CR pitch angle, $R = r_g / r_\mathrm{out}$ is the dimensionless rigidity, $\xi$ is the cosine of the angle between the wave vector and the mean field, $x$ is the wavelength of a mode normalised to $r_\mathrm{out}$, $k_\mathrm{max}$ is the maximum wavenumber of the turbulent cascade normalised to $1/r_\mathrm{out}$, $w = x R \sqrt{(1-\mu^2)(1-\xi^2)}$, $J_1$ is the Bessel function of the first kind of order 1, and $\Delta\mu = \sqrt{\mathcal{M}_A(1-\mu^2)}$ is the width of the resonance in pitch angle. For a medium where magnetic pressure greatly exceeds gas pressure and where the gas is collisional on scales of $r_g$, which is the case in the magnetotail, the cascade is terminated by viscous damping at a scale
\begin{equation}
    k_\mathrm{max} r_\mathrm{out} \approx \left(\frac{6 \mathcal{M}\mathcal{M}_A}{x_\mathrm{mfp}}\right)^{2/3}\left(1-\xi^2\right)^{-2/3},
\end{equation}
where $x_\mathrm{mfp}$ is the ratio of the particle mean free path to $r_\mathrm{out}$. Similarly, for transit-time damping the scattering rate is
\begin{eqnarray}
    K_\mu^\mathrm{TTD} & = & \frac{c}{r_\mathrm{out}} \frac{\sqrt{\pi}(1-\mu^2) \mathcal{M}_A^2}{2 R^2} \int_0^1 d\xi \int_1^{k_\mathrm{max} r_\mathrm{out}} dx\,
    \nonumber \\
    & &
     \frac{\xi}{x^{5/2}\Delta \mu}\left[J_1(w)\right]^2 \exp\left[-\frac{\left(\mu - \beta_\mathrm{A}/\xi\right)^2}{\Delta\mu^2}\right],
    \label{eq:KmuTTD}
\end{eqnarray}
where $\beta_\mathrm{A} = v_\mathrm{A}/c$. To evaluate these integrals numerically, we adopt the same choices for all parameters as in above; these in turn give $R =  8.2\times 10^{-6}$, $\beta_A = 6.6\times 10^{-4}$, $\mathcal{M} = 1$, $\mathcal{M}_A = 0.06$, and a Coulomb mean free path $\approx 10^{12}$ cm, so $x_\mathrm{mfp} = 1.0\times 10^{-7}$.

Using these choices to evaluate \autoref{eq:KmuG} and \autoref{eq:KmuTTD}, we find that as a function of $\mu$ the gyroresonant scattering rate reaches a maximum $K_\mu^\mathrm{G} \approx 1\times 10^{-11}$ s$^{-1}$ at $\mu\approx 0.7$. The value of $K_\mu^\mathrm{TTD}$ peaks extremely sharply at small $\mu$, but for $\mu > 0.7$, which is the range that is relevant for us since that is the range of initial pitch angles we find, we have $K_\mu^\mathrm{TTD} \lesssim 10^{-16}$ s$^{-1}$, so scattering due to transit-time damping is unimportant compared to gyroresonant scattering off fast modes. However, even this provides a scattering rate $\approx 1.5$ orders of magnitude smaller than the one required by the observations, and that we infer is provided by the streaming instability. This suggests that scattering due to an extrinsic cascade cannot explain our results.

However, we do caution that this conclusion is potentially sensitive to our assumed background gas density. If this were substantially higher that our assumed $1$ cm$^{-3}$, the Mach number would increase and the mean free path would decrease, thereby decreasing $x_\mathrm{mfp}$; both of these would in turn raise the gyroresonant scattering rate. We find that for densities $\gtrsim 10$ cm$^{-3}$, gyroresonant scattering does approach rates high enough to be competitive with streaming instability. We consider such high densities improbable, since even in the harsh environment of the Galactic Centre gas at such densities would likely cool and recombine quickly to become neutral, but we cannot fully rule out the possibility. Our results also depend somewhat less sensitively on our assumed velocity dispersion $\sigma_v$, which affects both $\mathcal{M}$ and $\mathcal{M}_A$, and if we were to increase this from 10 to 30 km s$^{-1}$, that would also be sufficient to render scattering by extrinsic turbulence comparable to streaming instability-induced scattering. However, we regard this possibility as unlikely because, as discussed in \sisecref{ssec:curvature}, there is no plausible mechanism to produce such large velocity dispersions in the tail; in any event, if the velocity dispersion were this large, that would likely be sufficient to bend the tail significantly, in which case no scattering is required to explain the displaced $\gamma$-ray emission in any event.\\

\subsection{Implications for CR self-confinement in the warm ionised ISM\\}
\label{ssec:wim}

As noted in the main text, our finding that CR self-confinement is effective for the CRe in Terzan 5 has implications for CR transport in the warm ionised medium (WIM) that fills a significant fraction of the Galactic volume away from the midplane. In order to do this, we must start by considering the ways in which the Terzan 5 environment differs from the typical WIM. One obvious difference is in CR number density: our MCMC fit for the CRe number density yields a 50th percentile value $n(>p_\mathrm{eq}) = 7.6 \times 10^{-13}$ cm$^{-3}$ with a 50th percentile value for $p_\mathrm{eq} = 2.5$ TeV. By contrast, for energies $\gg 1$ GeV, the CR proton number density in the Solar neighborhood is\cite{Gabici22a}
\begin{equation}
    n_\mathrm{CR,p}(>p) \approx 1.5\times 10^{-15} \left(\frac{p}{1\;\mathrm{TeV}/c}\right)^{-1.85}\mbox{ cm}^{-3}.
\end{equation}
Thus the electron number density at $\approx 2.5$ TeV$/c$ we have found in Terzan 5 corresponds roughly to the proton number density we would expect to find at $\approx 35$ GeV$/c$ in the WIM. We can therefore compare the growth and damping rates of the streaming instability in these two environments and for particles of comparable density. For damping, we again use \autoref{eq:gamma_turb}, since the WIM is similar to the Terzan 5 magnetotail in that both are likely magnetic pressure dominated ($\beta\lesssim 1$). For this purpose we adopt $n_\mathrm{H} = 0.1$ cm$^{-3}$, $B=6$ $\mu$G, and $r_\mathrm{out} = 100$ pc as typical WIM properties, and using these values we find that the damping rate in Terzan 5's magnetotail for $\sim$TeV electrons is likely a factor of $\sim 10$ larger than for $\sim 35$ GeV protons in the WIM, mostly due to Terzan 5's much larger magnetic field strength. For growth, note that the growth rate for CR protons is described by Equation 2 of the main text, simply with $m_e$ replaced by $m_p$. This means that, for both protons and electrons, we can express the growth rate of the streaming instability as
\begin{equation}
    \Gamma_\mathrm{si}(>p) = 2 e n_\mathrm{CR}(>p) \sqrt{\frac{\pi \mu_\mathrm{H}}{m_p^2 n_\mathrm{H}}} \frac{v_\mathrm{str}(>p)}{c}.
\end{equation}
During the initial stages of the streaming instability, before saturation is reached, $v_\mathrm{str}$ will be of order a few tenths of $c$ for both ISM protons and Terzan 5 electrons, so the growth rate in the WIM should be a factor of a few larger, driven by the somewhat lower density. 

Since we find that streaming instability appears to be effective in the Terzan 5 magnetotail, and that the growth rate is larger and the damping rate smaller in the WIM for CR protons at comparable density, this suggests that streaming instability should be effective at confining 35 GeV protons in the WIM as well. However, we do raise two cautions regarding this conclusion. The first is that we are considering only ionised gas, where turbulent damping dominates; the results might change considerably if neutrals are present and ion-neutral damping becomes important. Second, in the WIM it is possible that extrinsic turbulence suppresses the streaming instability by isotropising the CR population well enough that the streaming speed is below the Alfv\'en speed. We have shown above that this is likely not the case in Terzan 5, but we cannot rule out this possibility in all environments.\\

\subsection{Comparison with transport around other TeV sources\\}
\label{ssec:tev_halos}

As noted in the main text, our best-fit spatial diffusion coefficient is substantially larger than the values that some authors have inferred around other sources of TeV emission, the so-called ``TeV halos"\cite{Abeysekara17a, Lopez-Coto22a}. It is therefore interesting to explore the origin of this difference. We begin by noting that the claim of suppressed diffusion in TeV halos is not uncontroversial, and alternative explanations of the phenomenon have been advanced in the literature, including that diffusion is anisotropic and that in the TeV halos we are effectively measuring the slow, cross-field component rather than the fast, field-aligned one\cite{Liu19a} or that CRs in the TeV halos have not yet scattered sufficiently to have reached the diffusive regime at all\cite{Recchia21a}. However, if we neglect these possibilities and accept that diffusion rates are in fact suppressed in TeV halos to values smaller than we are finding for Terzan 5, this is nonetheless not surprising given the differences in the systems.

First we note that the main mechanism that we have found to be responsible for setting the diffusion coefficient around Terzan 5 -- streaming instability -- is not expected to operate at all in TeV halos given their vastly smaller CR luminosities\cite{Evoli18a}. Absent streaming instability, one possibility is that CRs in TeV halos are confined by extrinsic turbulence, and that the turbulent spectrum changes near TeV sources in ways that makes confinement more effective\cite{Lopez-Coto18a}. If this explanation is correct, then we would not expect it to apply to Terzan 5, since as noted above scattering by extrinsic turbulence in the Terzan 5 magnetotail is strongly suppressed by the enhanced magnetic field, which results in a very low Alfv\'en Mach number. This suppression is a result of the shock compression driven by Terzan 5's $\sim 100$ km s$^{-1}$ passage through the ISM, and will not be present around other TeV sources. 

A second possible explanation for suppressed diffusion in TeV halos is Bell's current-driven instability\cite{Bell04a}, which would induce rapid CR scattering\cite{Lopez-Coto22a}. However, the instability occurs only when the CRs carry a net current, and when the cosmic ray energy density is comparable to or larger than the magnetic energy density -- both conditions that are plausibly met in TeV halos, but not in Terzan 5. With regard to the former, while the winds of single pulsars plausibly contain an imbalanced flux of electrons and positrons, permitting a current to flow, the Terzan 5 wind is the aggregate of winds from $\sim 100$ pulsars, and thus it seems unlikely that a steady current can be set up. With regard to the latter, our best-fit magnetic field strengths and CR number densities for the Terzan 5 magnetotail imply a magnetic energy density that is $\sim 10 - 100$ times the CR energy density -- small enough to completely suppress Bell instability -- but again this is mainly due to Terzan 5's large velocity through the ISM, something that other TeV halos will not possess.

\end{si}

\bibliographystyle{naturemag}
\bibliography{main}

\begin{addendum}

\item[Code availability] The CR transport simulations in this paper were carried out with \textsc{criptic} \cite{Krumholz22a}, which is freely available from \url{https://bitbucket.org/krumholz/criptic/}. Calculation of the interpolation tables uses \textsc{cubature} \cite{cubature}, which is available from \url{https://github.com/stevengj/cubature}. The MCMC analysis uses \textsc{emcee} \cite{Foreman-Mackey13a}, which is available from \url{https://github.com/dfm/emcee}. The \textsc{Xspec} software that we use to calculate the X-ray absorption is available from \url{https://heasarc.gsfc.nasa.gov/xanadu/xspec/}. The input files used for the \textsc{criptic} simulations, along with source code to carry out all the analysis steps presented above, are available from \url{https://bitbucket.org/krumholz/terzan5}. All the software used in this project is available under an open source license.

\item[Data availability] The simulation data, interpolation tables, and MCMC chains are not included in the code repository due to their size ($\approx 200$ GB), but are available upon request from the corresponding author.

\item[Acknowledgements] RMC and MRK acknowledge 
support from the Australian Government through the Australian Research Council, award
DP230101055. This research was undertaken with the assistance of resources and services from the National Computational Infrastructure (NCI), award jh2, which is supported by the Australian Government. Part of this work as completed while MRK participated in a workshop at the Kavli Institute for Theoretical Physics (KITP), which is supported in part by grant NSF PHY-2309135.
PB acknowledges the financial support from the State Agency for Research of the Spanish Ministry of Science and Innovation under grants
PID2019-105510GB-C31/AEI/10.13039/501100011033, PID2019-104114RB-C33/AEI/10.13039/501100011033, and PID2022-138172NB-C43/AEI/10.13039/501100011033/ERDF/EU, and through the Unit of Excellence Mar\'{ı}a de Maeztu 2020–2023 award to the Institute of Cosmos Sciences (CEX2019-000918-M). PB acknowledges the financial support from Departament de Recerca i Universitats of Generalitat de Catalunya through
grant 2021SGR00679.
The authors thank 
Pasquale Blasi,
Matthew Baring, 
Maxim Barkov,
Holger Baumgardt, 
Geoff Bicknell,
Megan Donahue,
Enrico di Teodoro,
Miroslav Filopovic,
Oscar Macias,
Dougal Mackey, 
Emmanuel Moulin,
Madeleine McKenzie,
Ciaran O'Hare,
Barbara Olmi,
Brian Reville,
Gavin Rowell,  
Andreas Shalchi,
Deheng Song, and three anonymous referees
for useful communications.
RMC thanks Jim Hinton for alerting him to the peculiar TeV phenomenology of Terzan 5.

\item[Author contributions statement] RMC initiated the project and conceived the theoretical interpretation of the displaced TeV source. 
MRK conducted the numerical modelling.
The  text was written by MRK and RMC.
All authors were involved in the interpretation of the results and all reviewed the manuscript.

\item[Competing interests] The authors declare no competing interests.

\end{addendum}

\onecolumn

\setcounter{figure}{0}
\setcounter{table}{0}

\renewcommand{\figurename}{Extended Data Figure}
\renewcommand{\tablename}{Extended Data Table}

\begin{table}
    \centering
    \input{tables/mcmc_tab}
    \caption{
    \textbf{Marginal posteriors for model parameters.} From top to bottom, the parameters shown are the reduced $\chi^2$ value of the models, the pitch angle scattering rate $K_\mu$, the CR momentum for which the pitch angle scattering and synchrotron loss timescales are equal, the cutoff momentum in the CR injection spectrum, the index of the CR injection spectrum, the cosine of the maximum angle $\theta_0$ at which CRs are injected, the cosine of the angle $\theta_\mathrm{obs}$ between the magnetic field and our line of sight, the magnetic field strength, the total CR luminosity of Terzan 5, our rough estimate of the streaming instability-driven wave growth rate, and the synchrotron critical energy for CRs of momentum $p=p_\mathrm{eq}$, respectively. For each parameter, we report the value of the 2.28, 15.89, 50, 84.13, and 97.72 percentiles; the 50th percentile is the median, while the other percentiles correspond to the $\pm 1\sigma$ and $\pm 2\sigma$ limits.
    \label{tab:posteriors}
    }
\end{table}

\clearpage

\begin{figure}
    \centering
    \includegraphics{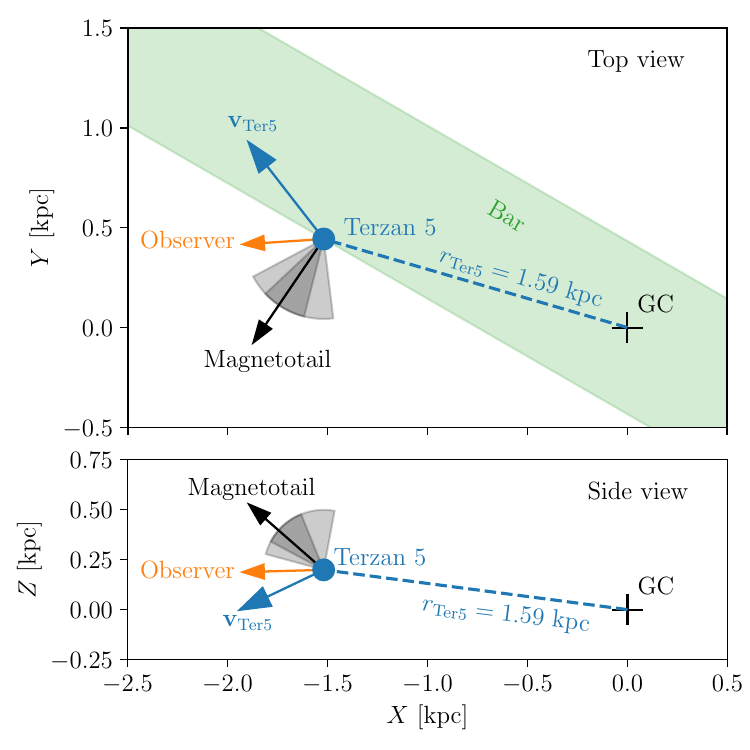}
    \caption{
    \textbf{Geometry of Terzan 5 relative to the Sun and the Milky Way.} We show Terzan 5's position (blue circle) and velocity (blue arrow) in a coordinate system where the Galactic Centre (black cross) lies at the origin and the Sun lies at $(X,Y,Z) = (-8.22,0,0.0208)$ kpc, in the direction indicated by the orange arrow. The top panel shows a top-down view of the Galactic Plane, and the bottom panel shows a side-on view. In both panels the dashed blue line shows the distance from the Galactic Centre to Terzan 5. The shaded green band in the top panel indicates the rough position of the Galactic Bar; we take the Bar thickness to be $\approx 1$ kpc and the orientation to be $30^\circ$ from the Sun-Galactic Centre line in the direction of Galactic rotation\cite{Bland-Hawthorn16b}. The black arrow indicates our 50th percentile value for the direction of Terzan 5's magnetotail as inferred from our 50th percentile value for $\mu_\mathrm{obs}$ together with the observed sky position of the TeV emission centroid\cite{Abramowski2011}; shaded grey arcs around this arrow show the $1\sigma$ and $2\sigma$ uncertainty intervals. We take the position, proper motion, and radial velocity of Terzan 5 from ref.~\cite{Baumgardt21a}, and we transform all quantities from sky coordinates to Galactocentric coordinates using the \textsc{astropy} Galactocentric coordinate package version 4.0\cite{Astropy-Collaboration13a, Astropy-Collaboration18a, Astropy-Collaboration22a}.
    \label{fig:T5geometry}
    }
\end{figure}

\begin{figure}
    \centering
    \includegraphics{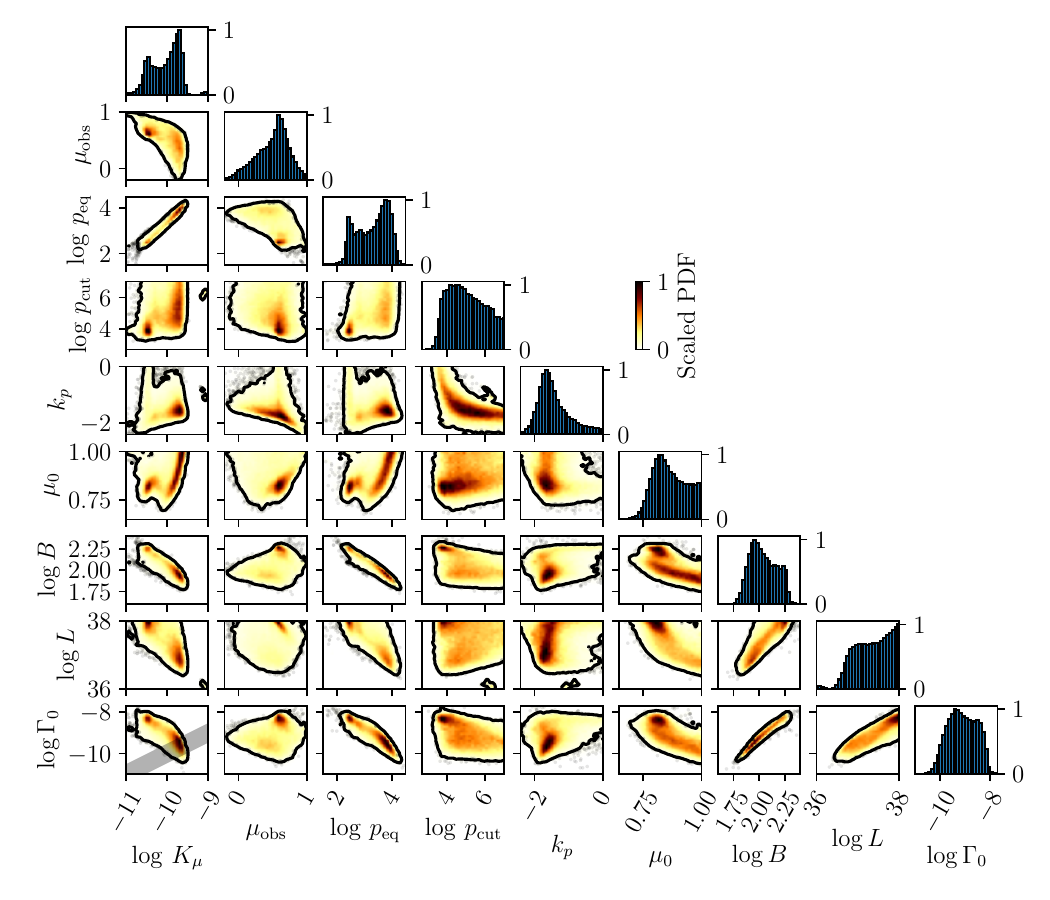}
    \caption{
    \textbf{Corner plot showing full posterior PDF determined by MCMC.} Panels along the diagonal show histograms of the marginal posterior PDF for each quantity, scaled to a maximum of unity. All other panels show the joint posterior PDF of two quantities; in these panels, colours show the 2D PDF scaled to a maximum of unity, and black contours enclose the marginal 95\% confidence interval on each pair of quantities. Points outside the contours show individual randomly-selected MCMC samples that fall outside the 95\% confidence interval. The quantities shown, and their units (omitted on the axis labels for reasons of space), are from left to right: log pitch angle scattering rate $K_\mu$ [s$^{-1}$], cosine of angle between magnetic field and line of sight $\mu_\mathrm{obs}$, log CR momentum for which loss and isotropisation times are equal $p_\mathrm{eq}$ [GeV$/c$], log CR momentum at which the injection distribution cuts off $p_\mathrm{cut}$ [GeV$/c$], injection spectral index $k_p$, cosine of the maximum injection angle $\mu_0$, log magnetic field strength $B$ [$\mu$G], log total CR kinetic luminosity $L$  [erg s$^{-1}$], and log streaming instability growth rate $\Gamma_0$ [s$^{-1}$]. In the lower left panel, the grey band shows the relation $K_\mu = \Gamma_0$ with a factor of 3 spread.
    \label{fig:corner}
    }
\end{figure}

\clearpage

\begin{figure}
    \centering
    \includegraphics{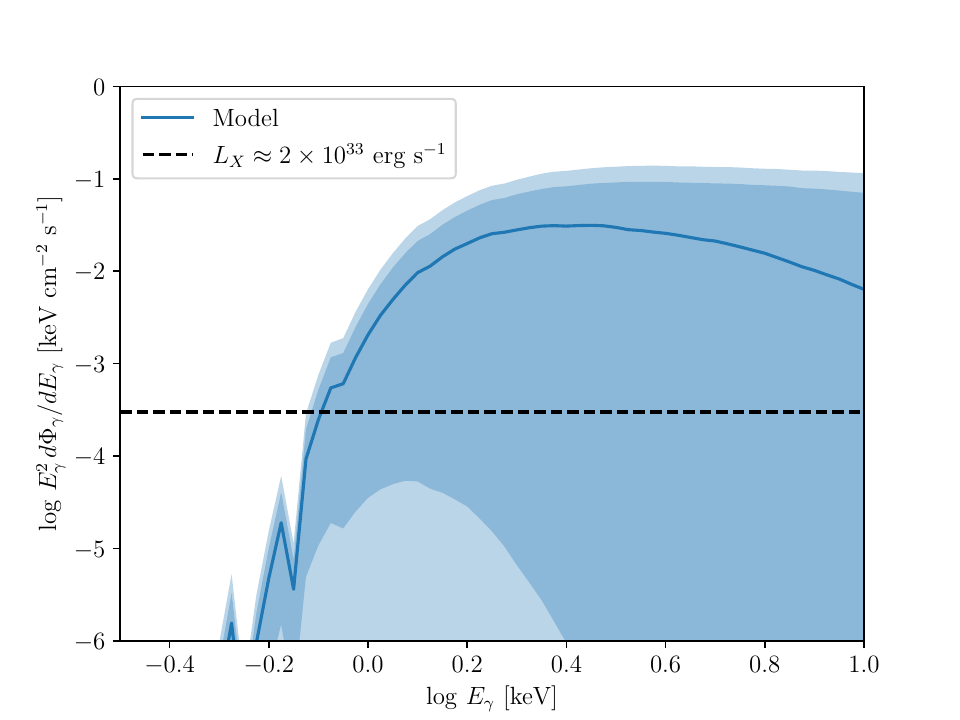}
    \caption{
    \textbf{Model-predicted synchrotron spectrum.} The blue line shows the median synchrotron spectrum as a function of photon energy $E_\gamma$ predicted by our best-fitting model, and the shaded blue bands around it show the 68\% and 95\% confidence intervals. The quantity shown includes the effects of interstellar absorption between Terzan 5 and the Sun, assuming a hydrogen column\cite{Eger10a} $N_\mathrm{H} = 2\times 10^{22}$ cm$^{-2}$; the sharp features visible below 1 keV correspond to absorption edges. The black dashed line is an approximate limit corresponding to $L_X = 4 \pi d_\mathrm{Ter5}^2 E_\gamma^2 (d\Phi_\gamma/dE_\gamma) = 2\times 10^{33}$ erg s$^{-1}$, the X-ray luminosity estimated by ref.~\cite{Eger10a}.
    }
    \label{fig:synchrotron}
\end{figure}

\end{document}

%% file: tables/mcmc_tab.tex
    \begin{tabular}{ll|rrrrr}
    \hline\hline
    \multirow{2}{*}{Quantity} & \multirow{2}{*}{Unit} & \multicolumn{5}{c}{Percentiles} \\ 
    & & 2.28 & 15.89 & 50 & 84.13 & 97.72 \\ \hline
   $\chi^2_\mathrm{red}$ & - & $1.71$ & $1.96$ & $2.14$ & $2.42$ & $2.89$ \\ 
   $\log K_\mu$ & s$^{-1}$ & $-10.66$ & $-10.44$ & $-9.97$ & $-9.70$ & $-9.59$ \\ 
   $\log p_\mathrm{eq}$ & GeV$/c$ & $2.34$ & $2.60$ & $3.40$ & $3.87$ & $4.14$ \\ 
   $\log p_\mathrm{cut}$ & GeV$/c$ & $3.62$ & $4.07$ & $5.00$ & $6.21$ & $6.88$ \\ 
   $k_p$ & - & $-2.11$ & $-1.81$ & $-1.51$ & $-0.95$ & $-0.22$ \\ 
   $\mu_0$ & - & $ 0.74$ & $ 0.80$ & $ 0.86$ & $ 0.95$ & $ 0.99$ \\ 
   $\mu_\mathrm{obs}$ & - & $-0.06$ & $ 0.24$ & $ 0.54$ & $ 0.71$ & $ 0.88$ \\ 
   $\log B$ & $\mu$G & $1.82$ & $1.91$ & $2.03$ & $2.19$ & $2.27$ \\ 
   $\log L$ & erg s$^{-1}$ & $36.59$ & $36.89$ & $37.39$ & $37.83$ & $37.98$ \\ 
   $\log \Gamma_0$ & s$^{-1}$ & $-10.18$ & $-9.71$ & $-9.14$ & $-8.51$ & $-8.20$ \\ 
   $\log E_\mathrm{sy}$ & keV & $-3.48$ & $-2.93$ & $-1.42$ & $-0.55$ & $-0.11$ \\ 
    \hline\hline
    \end{tabular}